\documentclass[11pt]{article}
\usepackage[left=2.5cm,right=2.5cm,top=3cm,bottom=3cm]{geometry}

\usepackage[english]{babel}
\usepackage{amsfonts,amsthm}
\usepackage{amssymb}
\usepackage{graphicx}
\usepackage[authoryear,round]{natbib}
\usepackage[normalem]{ulem}
\usepackage{algorithm}
\usepackage{algpseudocode}

\usepackage{float} 

\newcommand{\yobs}{y_{\text{obs}}}
\newcommand{\sobs}{s_{\text{obs}}}

\usepackage[textfont=it,font=small]{caption,subfig}
\usepackage{longtable}
\usepackage{rotating}
\usepackage{tablefootnote}
\usepackage{multirow}
\usepackage{array}

\usepackage{amsmath}
\usepackage{xr}
\makeatletter
\newcommand*{\addFileDependency}[1]{
  \typeout{(#1)}
  \@addtofilelist{#1}
  \IfFileExists{#1}{}{\typeout{No file #1.}}
}
\makeatother

\newcommand*{\myexternaldocument}[1]{%
    \externaldocument{#1}%
    \addFileDependency{#1.tex}%
    \addFileDependency{#1.aux}%
}

\myexternaldocument{supplementary}

\graphicspath{{/home/charlotte/Documents/Articles/En cours/calibration-paper/}}

\usepackage{authblk}

\usepackage{bbold}
\usepackage{color, colortbl}
\usepackage{enumitem}

\title{Calibration of a bumble bee foraging model using Approximate Bayesian Computation}
\date{}

\author[,1]{Charlotte Baey\thanks{Corresponding author: \texttt{charlotte.baey@univ-lille.fr}}}
\author[2,3]{Henrik G. Smith}
\author[2]{Maj Rundlöf}
\author[2]{Ola Olsson}
\author[3]{Yann Clough}
\author[3]{Ullrika Sahlin}

\affil[1]{Univ. Lille, CNRS, UMR 8524 - Laboratoire Paul Painlevé, F-59000 Lille, France}
\affil[2]{Lund University, Department of Biology, SE-223 62 Lund, Sweden}
\affil[3]{Lund University, Centre for Environmental and Climate Science, SE-223 62 Lund, Sweden}

\begin{document}

\maketitle

\begin{abstract}
1. Challenging calibration of complex models can be approached by using prior knowledge on the parameters. However, the natural choice of Bayesian inference can be computationally heavy when relying on Markov Chain Monte Carlo (MCMC) sampling. When the likelihood of the data is intractable, alternative Bayesian methods have been proposed. Approximate Bayesian Computation (ABC) only requires sampling from the data generative model, but may be problematic when the dimension of the data is high.

2. We studied alternative strategies to handle high dimensional data in ABC applied to the calibration of a spatially explicit foraging model for \textit{Bombus terrestris}. The first step consisted in building a set of summary statistics carrying enough biological meaning, i.e. as much as the original data, and then applying ABC on this set. Two ABC strategies, the use of regression adjustment leading to the production of ABC posterior samples, and the use of machine learning approaches to approximate ABC posterior quantiles, were compared with respect to coverage of model estimates and true parameter values. The comparison was made on simulated data as well as on data from two field studies.

3. Results from simulated data showed that some model parameters were easier to calibrate than others. Approaches based on random forests in general performed better on simulated data. They also performed well on field data, even though the posterior predictive distribution exhibited a higher variance. Nonlinear regression adjustment performed better than linear ones, and the classical ABC rejection algorithm performed badly.

4. ABC is an interesting and appealing approach for the calibration of complex models in biology, such as spatially explicit foraging models. However, while ABC methods are easy to implement, they often require considerable tuning. 

\end{abstract}



\section{Introduction}
Evidence of declines of pollinator populations \citep{ipbes2016} calls for accurate estimations of their status, spatial distribution and responses to future environmental change. Insect pollination is crucial for maintaining wild plant diversity as well as the production of many entomophilous crops \citep{Oller11,Gari13}, and bees play a major role in crops pollination \citep{Rader16}. In this context, spatially explicit foraging models accounting for bee mobility may serve the purpose of accounting for bee distribution in landscapes when estimating their population status, but also be used to generate predictions to support management and land-use decisions. Bee foraging in landscapes can be modelled based on diffusion from nests to floral resources \citep{Lons09,Hau17}, central place foraging theory \citep{OlssonBolin14,OlssonBolin15} or using agent-based modelling \citep{Becher14,Becher16}.
Calibrating these often complex and nonlinear models that produce high dimensional outputs is not straightforward. Parameters can be estimated based on literature or expert judgment, but confronting a model to field data is crucial to ensure its validity and ability to produce realistic predictions.
To this end, model calibration can be set up as an inverse modelling procedure to estimate model parameters by comparing model outputs with observations. In pattern-oriented modelling, summaries of generated model output are compared with corresponding summaries in observations. This method have previously been used to calibrate agent-based models of bees foraging \citep{Topping12,Becher14}. Inverse modelling have also been done to calibrate bee floral attractiveness and nesting densities in different land use classes \citep{Baey17,Gardner20}. Such statistical model calibration requires a probabilistic model for data given parameters, from which one can calculate the likelihood \citep{ohag01}. This generative model can be derived from a combination of observation and system processes \citep{Royle07}, where the system processes can be expressed by the spatially explicit foraging model. 

Parameter estimation using Bayesian inference allows incorporation of prior knowledge about the parameters and quantification of parameter uncertainty. Starting from a set of prior distributions, the aim is to compute the posterior distribution, i.e. the joint distribution of the parameters conditionally on the data \citep{Gelman95}. \textcolor{black}{Adopting a Bayesian point of view with informative priors can also guide the estimation process by providing regions of higher interest in the parameter space. In most cases, the posterior distribution is not available in a closed form and should be generated using sampling schemes such as Markov Chain Monte Carlo (MCMC) \citep{Tierney94,Robert04}.} 

\textcolor{black}{However, an additional issue may arise when dealing with complex ecological models. Indeed, these models are often defined as a set of hierarchical relationships involving latent variables which can be high dimensional. In this case, the likelihood of the model is obtained by integrating out the complete likelihood (i.e. the joint distribution of the data and the latent variables) over all possible values of the latent variables. This integration step is in most of the cases intractable. Therefore, classical MCMC approaches such as Metropolis-Hastings algorithm are unfeasible since they require evaluations of the likelihood function at each iteration. In this context, several alternatives have been proposed. When the complete likelihood is easy to compute, approaches which generate samples from the joint posterior distribution of the parameters and the latent variables can be used \citep{Wilson98}. However, they can prove to be very inefficient, for example if the dimension of the latent variables is too large. Another possibility is the Monte-Carlo within Metropolis (MCWM) algorithm, where the likelihood values required at each iteration are replaced by approximations based for example on importance sampling (IS) \citep{Oneill00,Beau03}. This approach can be computationally heavy and the high dimension of the latent variables space can hinder the efficiency of the algorithm. One can also resort to Variational Inference which has been recently extended in the context of intractable likelihoods by \cite{Tran17}, where exact evaluations of the likelihood are replaced by unbiased estimates. As with the MCWM approach, the computation of these estimates using Monte Carlo algorithms can be time consuming. From a frequentist point of view, intractable likelihoods issues can be handled using EM-type algorithms, even though they can be very difficult to set up in a high dimensional context, or using surrogate models for example, which are simpler versions of the original model carrying enough information about the parameters.}

\textcolor{black}{In this paper, we instead rely on approximate Bayesian computation (ABC).} Stemming from population genetics in the late 1990s \citep{Tav97,Beau02}, ABC has become a method of reference for highly complex models in a broad range of disciplines including biology \citep{Toni09}, ecology \citep{Beau10}, epidemiology \citep{Minter19} or economics \citep{Forn18}. \textcolor{black}{It has been successfully applied in the context of individual-based model in \cite{van2015calibration}, where it was also used as a tool to enhance model development. One of the many advantages of ABC is its flexibility, since the only requirement is to be able to simulate from the model. Moreover, these simulations can easily be performed in parallel, which is particularly relevant when dealing with complex models which can take a few seconds to run.} The basic idea of the algorithm is to generate several parameter values from given prior distributions, and to compute simulated values using the sampled parameters and the data generative model \citep{Csi10}. Then, only those parameter values leading to simulated data which are close enough to the observed data are retained.  Several generalizations and extensions of the algorithm have been proposed to handle issues that may arise in practice, such as the high dimension of the data and the choice of a criterion to measure the distance between simulated and observed data.
%

In this paper, we give an overview of different ABC methods usable \textcolor{black}{for the calibration of complex models with highly dimensional and often noisy observations}, and compare their performances on a spatially explicit bumble bee foraging model. This is a deterministic model based on central place foraging theory \citep{OlssonBolin14} combined with a probabilistic model for the field observations, and described in Section \ref{sec:model}.
Comparison of the ABC methods is first made on a set of simulated data, and then applied to field data from two field studies on pollinator abundance in southern Sweden. Parameter estimation is described in Section \ref{sec:abc} and calibration performances of the different algorithms are evaluated on their abilities to accurately estimate the model parameters, based on simulated data generated under the model and on field data (Section \ref{sec:results}).


\section{Material and methods}
\subsection{The Central Place Foraging model}\label{sec:model}

Here, we briefly describe the central place foraging (CPF) model used (for details see \cite{OlssonBolin14} and \cite{OlssonBolin15}). It is built on the assumption that fitness-maximizing animals (i.e. bumble bees) nest in a central place to which they collect food in the surrounding landscape. 
Since commuting between the nest and foraging patches requires time and energy, bees are willing to fly to a distant patch if and only if it provides enough food of suitable quality, while at shorter distances also lower quality patches are visited. The model requires two types of inputs: a rasterised map $\mathcal{M}$ giving the land-use category of each pixel (e.g. grassland, urban area, woodland, ...), and a set of parameters which we denote by $\theta$.
The model used is a modified version of the original CPF model. It is based on the CPF foraging algorithm \cite{OlssonBolin15}, but we replaced the equation for the maximum distance a forager from a given nest is prepared to fly (equation \eqref{eq:traveldist}). Following \cite{Lons09}, we did not explicitly include population growth across the season, but ran the model using  season-dependent inputs across three sequential seasonal periods.
For each seasonal period, we assigned floral and nesting values to each land-use category, that reflects the attractiveness and quantity of floral resources it provides, as well as the attractiveness in terms of nesting. Both floral and nesting values are recorded on a 0-1 scale, with 0 representing no attractiveness and 1 maximum attractiveness of the given land-use category. Floral maps were generated from land use maps (see below) by sampling random floral values from parameters calibrated on expert judgment or in some cases data \citep{Baey17}.


\paragraph{General behaviour.}
We define the minimum floral value resulting in any visit by bees as $f_0$, and the maximum distance travelled by a bee as $\tau_0$.
Now, for a patch of floral quality $f$, with $f\geq f_0$, the maximum distance an individual bee is prepared to fly to reach it is given by:
\begin{equation}
\tau_f = \tau_0 \left(1 -\frac{f_0}{f}\right).
\end{equation}
I.e., $\tau_0$ is the maximum distance a bee is willing to fly for a patch of infinite floral quality.

A bee nesting in patch $i$ will visit patch $j$, if its floral quality $f_j$ is high enough with respect to the distance between the two patches. We define by $\Delta_{ij}$ the difference between the maximum distance the bee is willing to fly for a patch of floral quality $f_j$ and the actual distance $d_{ij}$ between patches $i$ and $j$:
\begin{equation}
\Delta_{ij} =  \tau_0 \left(1 -\frac{f_0}{f_j}\right) - d_{ij},
\end{equation}
This quantity will be largest for a patch of ``infinite'' floral quality located adjacent to the nest. In other words, $\Delta_{ij}$ is a measure of the distance bees nesting in patch $i$ will spare by flying to patch $j$ compared to how long they were willing to fly for a patch of that quality.
Then, the suitability of a nest in patch $i$ is defined by:
\begin{equation}
    s_i = \sum_{j} \Delta_{ij} \mathbf{1}_{\Delta_{ij} > 0},
\end{equation}
where the sum is over all the patches in the landscape. This quantity can be viewed as measuring the distance a bee will spare flying when its nest is surrounded by enough patches of good quality. The more patches with high floral quality that are located around the nest, the higher the suitability of the nest.

\begin{figure}
	\centering
	\includegraphics[scale=0.75]{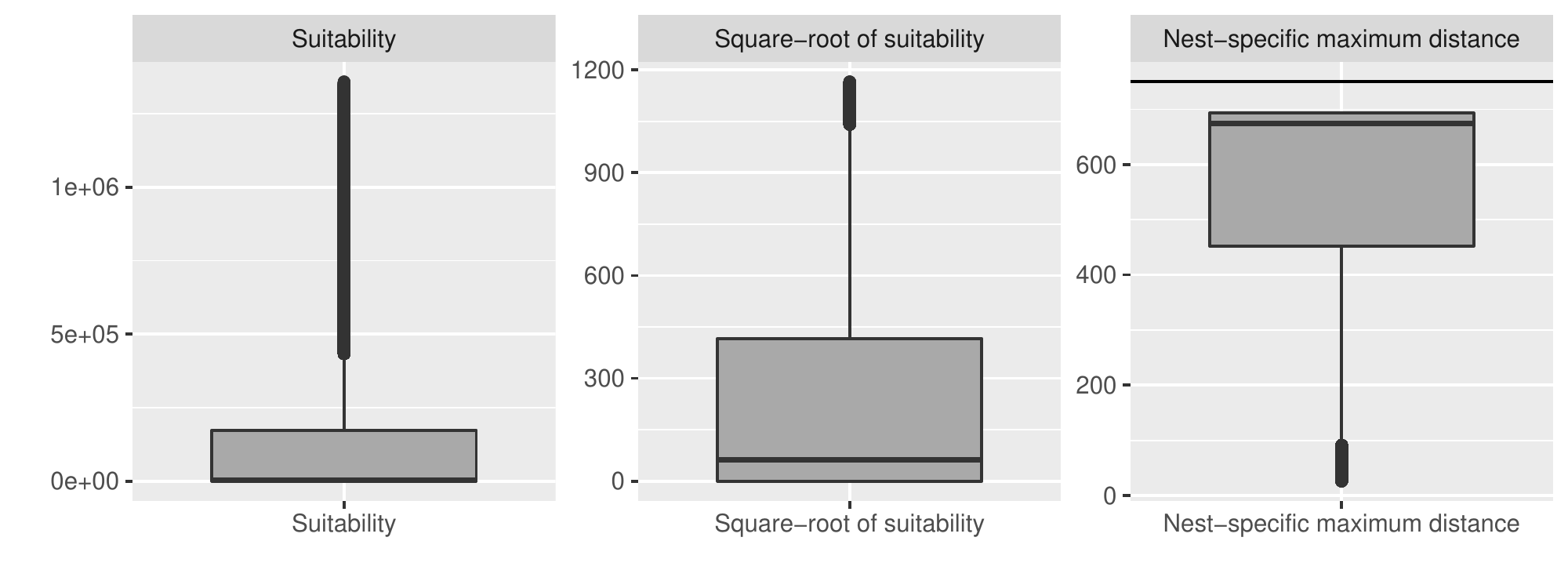}
	\caption{Example of the range of variation of nest suitability values, of its squared root and of the resulting nest-specific $\tau_i$ for $\tau_0$ = 750m, $a$=500 and $b=200$.}
	\label{fig:nestspecific}
\end{figure}

\paragraph{Optimization of foraging.} 
An individual bee nesting in patch $i$ seeks to optimize where to forage by exploiting surrounding patches according to preferences determined by a trade-off between distance and floral quality. This means that a bee with a nest surrounded by patches of flowers of high suitability will exploit fewer patches further away compared to a bee who’s nest surrounded by patches of low suitability.  This lead to the definition of a new ``nest-specific" maximum distance the bee is prepared to travel from its nest in patch $i$:
%
\begin{equation}\label{eq:traveldist}
\tau_{i} = \frac{\tau_{0}}{1 + \exp((\sqrt{s_i}-a)/b)}.
\end{equation}
The definition of this ``nest-specific'' maximum distance allows bees to adapt their behavior to account for differences in landscape structure. \textcolor{black}{We chose a logistic curve to enhance the interpretation of the parameters: $a$ is the inflexion point i.e. the suitability value for which the nest-specific maximum distance is equal to half the maximum $\tau_0$ and $b$ is the slope of the logistic curve. Parameter $b$ is positive, so that $\tau_i$ is a decreasing function of $s_i$: the higher the suitability of their nest, the closer to the nest the bees will fly. Since suitability values are computed as sums over all the patches in the landscape of quantities varying from 0 to $\tau_0$, they can be very high,  we used a squared root scale in the logistic function \eqref{eq:traveldist} (see Figure \ref{fig:nestspecific} for an example of the ranges of variation of $s_i$, $\sqrt{s_i}$ and $\tau_i$).}

Similarly to $\Delta_{ij}$, we define a new quantity $\Delta^*_{ij}$ using the nest-specific maximum distance $\tau_i$:
\begin{equation}
\Delta^*_{ij} = \tau_{i} \left(1 - \frac{f_0}{f_j}\right) - d_{ij},
\end{equation}
$\Delta^*_{ij}$ can be seen as the contribution from patch $j$ to fitness of the bees nesting in patch $i$. 

Then, the rate of foraging bees from a nest in patch $i$ to floral resources in patch $j$ is set to:
\begin{equation}
r_{i \rightarrow j}  = q_i \frac{\Delta^*_{ij}}{\sum_{j=1}^J \Delta^*_{ij}}.
\end{equation}
where $q_i$ is the nesting value.

Finally, the intensity of (instantaneous) overall rate of bees visiting patch $i$ is then defined as the sum of foraging rates by:
\begin{equation}\label{eq:vrcpf}
    \nu_i (\theta , \mathcal{M}) = \sum_{j=1}^J r_{j \rightarrow i},
\end{equation}
where $\theta = (\tau_0,f_0,a,b)$ is the vector of parameters from the CPF model \textcolor{black}{(see also Table \ref{tab:params})}, and $\mathcal{M}$ is the (fixed) map used as an input to the CPF model, containing informing about the landscape structure and land-use of each cell in the rasterized landscape.



\subsection{Data}\label{sec:data}
Observations of bees abundances are extracted from two studies monitoring pollinator abundances in southern Scania, thereafter called respectively STEP and COST. In this study, we focus on bees from the \textit{Bombus terrestris} species. \textcolor{black}{A total of 790 measurements of bumble bees abundances are available from these two studies, covering four different years and up to three periods along the season.}



\begin{figure}[h]
\centering
\includegraphics[scale=0.4]{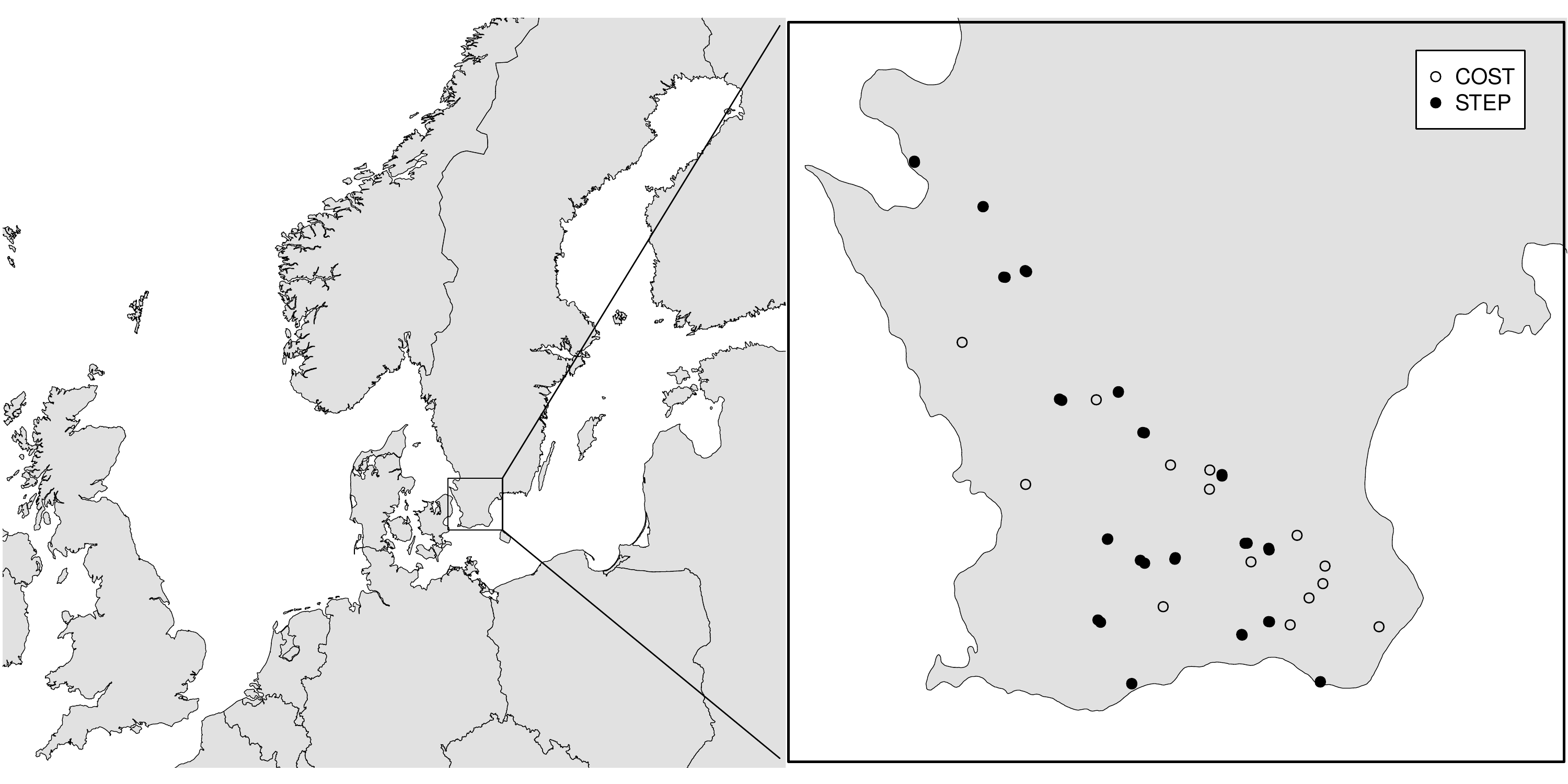}
\caption{Sampling locations from the STEP (16 locations) and COST (19 farms) studies.}
\label{fig:step}
\end{figure}

\paragraph{STEP data.}
We use a Swedish dataset from the European project \emph{Status and Trend of European Pollinators} (STEP), which took place in different European countries including Sweden in 2011 and 2012 \citep{Holzs16}.  \textcolor{black}{This study surveyed several pollinators among which honey bees, bumble bees, hoverflies and wild bees. In this paper, we focus only on bumble bees from the \textit{Bombus terrestris} species.} 
Data were collected in 16 different locations in Scania in southernmost Sweden (Figure \ref{fig:step}).
Each location consisted of three sites within a circle of 2 km radius, each corresponding to a specific land use category: oilseed rape field, semi-natural grassland and field edge.
In 2012, a wildflower strip was also surveyed in 8 locations. 
%
In each site and year, the number of bumble bees was recorded at two occasions for each floral period considered (early and late in the season). Bumble bee numbers were recorded along 150m\textup{2} transects (150m$\times$1m) during 15 minutes. Foraging and flying bees were counted separately.
\textcolor{black}{A total of 513 bumble bees abundances are available.}

\paragraph{COST data.}
In the COST project, data was collected around 19 farms in Scania, with three habitat types surveyed at each farm: cereal field, ley field, and semi-natural grassland \citep{Carrie18}. \textcolor{black}{This study surveyed bumble bees and butterflies, but we focus on bumble bees from \textit{Bombus terrestris} species in this paper.}

Bumble bees were surveyed in 2016 and 2017, at different occasions covering the two periods represented in the STEP study plus an additional period later in the season.
The number of bumble bees was recorded along a 200m\textup{2} transect (100m$\times$2m) for a period of 10 min. No distinction was made between foraging and flying bees.
\textcolor{black}{A total of 278 bumble bees abundances are available.}



%
%

\paragraph{Land use maps.}
Information about land use was extracted from the Swedish National Land Cover Database from the Swedish Environmental Protection Agency (based on satellite data with 10m resolution in combination with other layers) and the Swedish Integrated Administration and Control System which is a geographical database on farmland use in Sweden organized by the Swedish Board of Agriculture and maintained to administrate agricultural subsidies and agri-environment schemes. The latter was used to provide more detailed information on land use within agricultural land. 
\textcolor{black}{The choice of the spatial resolution was made as a compromise between the computation time (which rises as the resolution increases) and the accuracy of the landscape description. At a larger resolution scale, difficulties could appear when merging different landuse types, for example on the definition of the resulting landuse type for the merged cell. For this resolution scale, running the model on all the landscapes took approximately 1 hour and 45 minutes on a 4 cores Intel(R) Core(TM) i3-6100T CPU @ 3.20GHz processor.}

\subsection{Bayesian formulation of the model}
Parameter estimation is conducted in a Bayesian framework. In this section we describe the likelihood of the data and the priors for the parameters.
Let $y_{ijk}, i=1, \dots, n, j = 1, \dots, J, k=1,\dots,K$ denote the observations of the number of bees on site $i$, year $j$ and period $k$. 
Each sampling site is associated with a specific study. Each study, in turn, is associated with a given set of landscapes and was conducted during different years, with no overlap between studies. To reduce computation time, the model is not run on the whole map of Scania, but on a set of smaller landscapes. For each study, a 10$\times$10 km$^2$ landscape centered on each surveyed oilseed rape or cereal field was used. These covered the sampling sites mentioned in section \ref{sec:data}, corresponding to oilseed rape fields, field borders and semi-natural grassland for the STEP study, and to cereal field, ley field and semi-natural grassland for the COST study. We assume that conditionally on the landscape structure and on the model, the observation made at different locations are independent.

\paragraph{Likelihood.}
We denote by $\lambda_{ijk}$ the real intensity of the visitation rates process on sampling site $i$, year $j$ and  period $k$. The data generative model is specified as the following hierarchical model:
\begin{itemize}
	\item \textbf{part 1:} observed bee abundance varies according to a Poisson distribution with an intensity depending on site, year and period:
	 \begin{align}\label{eq:obs}
    y_{ijk} \mid \lambda_{ijk} , \theta & \sim \text{Poi}(c_{i} \cdot \lambda_{ijk}), 
\end{align}
where $c_{i}$ is a known scaling parameter accounting for the time window of the observation process and the area of the sampling site. More specifically, $c_{i} = d_i \cdot a_i$, with $d_i$ and $a_i$ respectively the duration of observation process and the area of the sampling site $i$.

	\item \textbf{part 2:} the realised (log) intensity of the Poisson distribution on a site at a given time can be characterised as normally distributed with a mean given by the CPF model and a time period-specific parameter:
\begin{align}
\label{eq:logint}
    \log \lambda_{ijk} & = \log \nu_i(\theta, \mathcal{M}_{ijk}) + \beta_1 + \sum_{l=2}^K \beta_{l} \mathbb{1}_{l=k} + \varepsilon_{ijk}, \quad \varepsilon_{ijk} \sim \mathcal{N}(0,\sigma^2).
\end{align}
Since there are differences in population sizes at landscape scale between periods within a year, not considered in the CPF model, period-specific parameters are introduced: $\beta_1$ is the baseline effect on period 1, and $\beta_l$, for $l=2,\dots,K$ correspond to the development of population size compared to period 1 (i.e. the effect of period $k$ on the log intensity is $\beta_1 + \beta_k$)
\end{itemize}


\begin{table}
\centering
\caption{\textcolor{black}{Summary of model parameters and observation parameters}}
\label{tab:params}
\textcolor{black}{\begin{tabular}{p{2cm}cp{11cm}}
\hline
& Parameter & Interpretation \\
\hline
\multirow{4}{2cm}{\centering Model parameters} & $\tau_0$ & maximum distance a bee is prepared to fly for a patch of infinite floral quality, i.e. asymptotic value for the nest-specific maximum flying distance \\
 & $f_0$ & lowest floral quality a bee will ever be visiting \\
  & $a$ & nest suitability value (in a square root scale) resulting in a nest-specific maximum distance equals to half the maximum distance $\tau_0$ \\
  & $b$ & slope of the nest-specific maximum distance curve\\
  \hline
\multirow{2}{2cm}{\centering Observation parameters} & $\beta_k$ & period-specific scaling parameter for the population size\\
 & $\sigma^2$ & observation noise \\
\hline\end{tabular}}
\end{table}

The complete vector of parameters is given by $\psi = (\theta, \omega)$, where $\theta$ is the aforementioned vector of parameters from the CPF model \textcolor{black}{which are needed to compute the visitation rate $\nu_i$ in equation \eqref{eq:logint}}, and $\omega = (\beta_1, \dots, \beta_K, \sigma^2)$ is the set of parameters corresponding to the observation process, \textcolor{black}{i.e. the parameters linking the visitation rate given by the CPF model and the mean intensity of the Poisson distribution in equation \eqref{eq:obs}}. \textcolor{black}{A summary of all the parameters is given in Table \ref{tab:params}.} Our main objective is to estimate $\theta$, and $\omega$ can therefore be seen as nuisance parameters. We denote by $\yobs$ the vector of observations, and by $p$ the total number of parameters to be calibrated (i.e. the dimension of $\psi$). 
\textcolor{black}{Combining equations \eqref{eq:obs} and \eqref{eq:logint} we can define the likelihood of the data as:}
\begin{align}\label{eq:likelihood}
\begin{split}
L(\yobs \mid \psi) & = \prod_{ijk} L(y_{ijk} \mid \psi) \\
	& = \prod_{ijk} \int p(y_{ijk} , \lambda\textcolor{black}{_{ijk}} , \psi) \ d\lambda \\
	& = \prod_{ijk} \int p(y_{ijk} \mid \lambda\textcolor{black}{_{ijk}} , \psi) \ p(\lambda\textcolor{black}{_{ijk}} \mid \psi) \ d\lambda\textcolor{black}{_{ijk}} \\
	& = \prod_{ijk} \frac{1}{\sqrt{2 \pi} \sigma y_{ijk}!} \int_0^{+\infty} e^{-\lambda\textcolor{black}{_{ijk}}} \lambda\textcolor{black}{_{ijk}}^{y_{ijk}-1} \exp\left( - \frac{(\log \lambda\textcolor{black}{_{ijk}} - \log \nu_i(\theta,\mathcal{M}_{ijk}) - \beta_k)^2}{2 \sigma^2} \right) \ d\lambda\textcolor{black}{_{ijk}}
\end{split}
\end{align}

This Poisson-lognormal distribution \citep{Izs08} is commonly used to model  \textcolor{black}{count data (see \cite{Bul74} in the context of ecological data, or \cite{Winkel08} in the context of econometric data). Indeed, simple Poisson distributions are often not appropriate to model over-dispersed data especially when there is an excess of 0s. A classical alternative in this case is to use the negative binomial distribution, which can also be written as a hierarchical model where, in the first stage, observations are modeled as in equation \eqref{eq:obs} and, in the second stage, the intensity of the Poisson distribution is assumed to be Gamma distributed. In the Poisson-lognormal model, the use of a Gaussian distribution in the second stage allows for more flexibility and an easier interpretation of the mean intensity.
However, the integral appearing in equation \eqref{eq:likelihood} is intractable and cannot be computed analytically. This prevents the use of classical methods such as MCMC algorithms which require the evaluation of the likelihood function. More details are given in section \ref{sec:abc}.}

\paragraph{Prior distributions.} 
The prior for parameters $\tau_0$ and $f_0$ is specified with some degree of precision using informal expert judgment \citep{Hau17}. For example, the upper bound for $\tau_0$ was set to be 1000 m based on previous results showing that the majority of foraging occurs within that range \citep{Osborne08}. Log-normal priors are chosen for $\tau_0$ and $f_0$ as these are non-negative numbers. Flat priors,  but within realistic ranges, are used for the other parameters. 
\begin{align}\label{eq:priors}
\begin{split}
\tau_0 & \sim \mathcal{LN}_{[0,1000]}(\log (1000),1) \\
 f_0 & \sim \mathcal{LN}(\log(0.1),1) \\
  a & \sim \mathcal{U}([100,1000]) \\
   b  &\sim \mathcal{U}([100,1000]) \\
\beta_k & \sim \mathcal{N}(0,100), \quad k = 1,\dots,K \\
 \sigma^2 &\sim \mathcal{IG} (1,1) 
 \end{split}
\end{align}
%
Assuming independence between the prior distributions, the joint prior distribution for parameter $\psi$ is given by:
\begin{align*}
\pi(\psi) = \pi(\tau_0) \pi(f_0) \pi(a) \pi(b) \left(\prod_{k=1}^K \pi(\beta_k) \right) \pi(\sigma^2)
\end{align*}


\paragraph{Posterior distribution.} The posterior distribution of the parameters is then defined as:
\begin{equation}\label{eq:post}
\pi(\psi \mid \yobs) \propto L(\yobs \mid \psi)  \ \pi (\psi) = \prod_{i,j,k} L(y_{ijk} \mid \psi) \ \pi(\psi).
\end{equation}

\subsection{Calibration using Approximate Bayesian Computation (ABC)}\label{sec:abc}

\textcolor{black}{Due to the aforementioned issue of likelihood intractability, c}alibration of the model is made using an ABC approach, where the computation of the likelihood is replaced by the generation of samples from the model. 
\textcolor{black}{Starting with a threshold $\varepsilon$ and a distance $d$ on the set of observations, the first and simplest version of the ABC algorithm is the ABC rejection sampling}:
\begin{enumerate}
  	 	\item draw samples $\psi^{(m)} = (\theta^{(m)},\omega^{(m)}), \ m=1,\dots, M,$ from the prior distribution
		\item generate the associated sets of observations $y^{(m)}, \ m=1,\dots,M$ using equations \eqref{eq:logint} and \eqref{eq:obs}
		\item for $m=1,\dots,M$,  keep sample $\psi^{(m)}$ if $d(\yobs,y^{(m)}) \leq \varepsilon$ 
\end{enumerate}

%

\textcolor{black}{As a result of this algorithm we get a sample of size $M_\varepsilon$, with all the accepted sets of parameters, each of them following the ABC posterior distribution. 
The approximation of the posterior distribution is better when $\varepsilon$ is small, and it can be shown that the ABC posterior converges to the true posterior when $\varepsilon$ tends to 0, and to the prior when $\varepsilon$ tends to infinity. }
%
However, due to the curse of dimensionality, the distance between any simulated $y^{(m)}$ and $\yobs$ tends to be arbitrarily large when the dimension of the data increases. Therefore, one has to either increase dramatically the number of ABC iterations $M$ or the threshold $\varepsilon$ to maintain a reasonable value for the final number of accepted values $M_\varepsilon$. In the former case, computation time can be burdensome, and in the latter case the quality of the results is degraded.

Several extensions have been proposed to circumvent these issues. A first suggestion is to consider a smoothing kernel $K$ instead of a crude rejection, i.e. each sample $y^{(m)}$ is associated to a weight proportional to $K(d(\yobs,y^{(m)}))$. \textcolor{black}{In this case, all the samples are used, which reduces the waste of computation time.}
These weights can then be used in an importance sampling scheme to compute the ABC posterior distribution. 
%
%
\textcolor{black}{To deal with the high dimension of the observed data, a common practice is to work with a set of \textit{summary statistics}, i.e. a function $s(\cdot)$ of lower dimension than the observed data. The choice of the summary statistics is crucial: they should be informative enough about the underlying biological processes, i.e. carry as much information as the original data, while being less noisy than the vector of original data, and possibly reducing the dimension of the data. A simple example in the case of a normally distributed sample of size $n$ is to define $s(\cdot)$ as a 2-dimensional vector containing the sample mean and the sample variance. A bad choice of the summary statistics can lead to a large loss of information. For example, the summary statistics computed on two different datasets can turn out to be identical, complicating the task of the ABC algorithm based on these summary statistics.
Most of the time, the main difficulty is not to find summary statistics \textit{per se}, but to find those that are relevant from a biological point of view. An interesting strategy could be to first identify a set of statistics which can be large and then deal with this high dimensionality with appropriate methods. To this end, several methods have been suggested that either select subsets of relevant statistics or are able to deal with high dimensional statistics.} 

\begin{figure}
\centering
\includegraphics[scale=1]{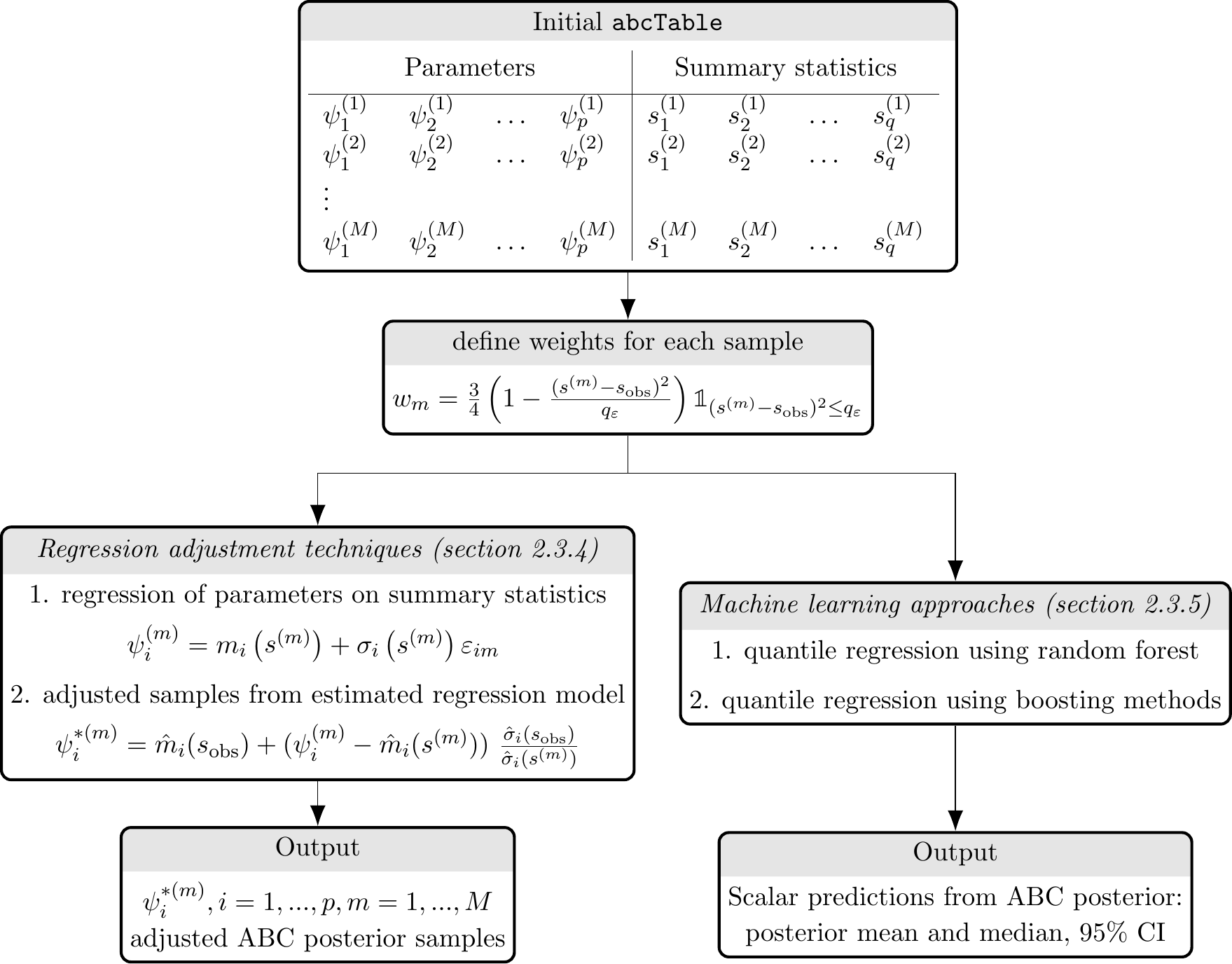}
	\caption{Summary of the different types of methods considered in the paper.}
	\label{fig:abccat}
\end{figure}

\textcolor{black}{Our analysis is thus divided into the following steps (see also figure \ref{fig:abccat}): i) we define a first set of summary statistics in collaboration with experts from the ecological field, ii) we sample parameter values from the priors and compute the associated summary statistics and kernel weights, iii) we compare different approaches building upon these summary statistics to either approximate the posterior distribution of the summary statistics or estimate key quantities from this posterior distribution.}

\textcolor{black}{Weights were assigned to each simulated parameter values using an Epanechnikov kernel and the summary statistics were scaled so that only the $\varepsilon M$ points which are the closest to the observed summary statistics have a positive weight, and we compared two different threshold values $\varepsilon=2.5\%$ or $\varepsilon=5\%$.
Table \ref{tab:summet} summarizes the different methods that were compared in this paper. The code and the data are available in the git repository \texttt{https://github.com/baeyc/bloomcpf}. }

\begin{table}
\centering
\small{
\caption{Summary of the different methods used in the paper}
\label{tab:summet}
\begin{tabular}{lp{8cm}lc}
\hline
Name & Description & Reference & \texttt{R} package \\
\hline
Rej & ABC with rejection sampling & \cite{Tav97} & \texttt{abc} \\
\hline
LocLH & Adjusted ABC samples from local linear heteroscedastic model & \cite{Beau02} & \texttt{abc} \\
LocNLH & Adjusted ABC samples from nonlinear heteroscedastic model & \cite{Blum10} & \texttt{abc} \\
ANLH & Adjusted ABC samples from adaptive nonlinear heteroscedastic model & \cite{Blum10} & \texttt{abc}, \texttt{e1071}\\
RFA & Adjusted ABC samples from nonlinear regression via random forests & \cite{Bi2022} &  \texttt{ranger} \\
\hline 
wqRF & Quantile regression via random forests (weighted samples) & \cite{Rayn19} & \texttt{abcrf} \\
uwqRF & Quantile regression via random forests (unweighted samples) & \cite{Rayn19} & \texttt{abcrf} \\
qGBM L1 & Quantile regression via gradient boosting ($L_1$ loss)& & \texttt{gbm}\\
qGBM L2 & Quantile regression via gradient boosting ($L_2$ loss) & & \texttt{gbm}\\
\hline
\end{tabular}
}
\end{table}

\subsubsection{Initial choice of summary statistics.}\label{sec:sumstat}


\textcolor{black}{In this section, we describe the first set of summary statistics that was defined in association with ecological experts. This initial set of summary statistics need not be too informative since specific methods will be used in the sequel to deal with summary statistics which do not carry enough information.}

Here, summary statistics were defined as the interquartile range and the number of 0's observed per sampling site, per period and per year, but combining all types of habitat (which resulted in 210 summary statistics) and the interquartile range and number of 0's observed per habitat type, per period and per year, but combining all sampling sites (which resulted in 194 summary statistics). It allowed for a first reduction of the dimension, from 790 observations to 404 summary statistics.

The division per site, period, year and habitat was made to capture characteristics attributed to these groups\textcolor{black}{: aggregation across habitats allows to account for the differences in population sizes between landscapes, whereas habitat-specific summaries allow to capture the joint effect of population size and relative attractiveness of the habitats}. These summary statistics were chosen to characterize the variability of the observations within the different groups, and to consider the high frequency of 0's. \textcolor{black}{The high number of zeros in the original data lead us to consider measures that are robust to outliers and to the presence of these zeros. For this reason, we chose the interquartile range to measure the variability of the data, instead of the standard deviation.}

\subsubsection{First point of view: approximation of the ABC posterior of the summary statistics}\label{sec:dimred}

In this section, we \textcolor{black}{consider the first point of view of producing samples from the ABC posterior distribution of the summary statistics and discuss different approaches based on regression adjustment.}
The main idea behind these approaches is to build a relationship between the parameter values and the summary statistics values, usually through regression techniques, and to use this regression layer to produce adjusted samples from the ABC posterior distribution of the summary statistics.
The general model is given by \citep{Blum10} :
\begin{equation}\label{eq:regmod}
\psi_i^{(m)} = m_i \left(s^{(m)} \right) + \sigma_i \left(s^{(m)}\right)\varepsilon_{im}, \quad i=1,\dots, p
\end{equation}
with $\varepsilon_{im}$ a set of iid zero-mean random variables, and where function $\sigma_i$ allows to account for heteroscedasticity.

\textcolor{black}{We compared different approaches. First, we considered the }
linear homoscedastic case \textcolor{black}{(i.e. $\sigma_i(s^{(m)}) = 1$)}, \textcolor{black}{where} estimation of $m_i$ is performed \textcolor{black}{using} minimum weighted least squares. 
%
%
\textcolor{black}{Then, we considered} the nonlinear and heteroscedastic case, \textcolor{black}{where} $m_i$ is estimated using feed-forward neural network, while estimation of $\sigma_i$ is performed using a second regression model for the log of the squared residuals. \cite{Blum10} proposed an adaptive two-stage version of this method: after a first step where adjusted sampled values are obtained via equation \eqref{eq:summstatnch}, in a second step the support of the ABC posterior distribution is estimated from this first set, e.g. using support vector machines. Then, a new nonlinear heteroscedastic regression model is built on the adjusted samples falling inside the estimated ABC posterior density support.
\textcolor{black}{Finally, based on a recent paper from \cite{Bi2022}, we considered the homoscedastic nonlinear case where $m_i$ is estimated using random forests (RF), partly because of their robustness and their ability to select the most relevant variables from a set of potentially large explanatory variables. In their paper, they proved in particular that the mean computed on the adjusted ABC posterior sample is an unbiased estimator of the ABC posterior mean.
RF is a method introduced by \cite{Brei01} based on the aggregation of several regression trees, each of them being built on a bootstrap sample of the data, and using only a random subset of all the available explanatory variables. The RF results in a partition of the space of explanatory variables, and in a piece-wise constant prediction on each set of this partition.  }
%
After estimation of $m_i$ and $\sigma_i$, adjusted samples from the ABC posterior distribution can be obtained via:
\begin{equation}\label{eq:summstatnch}
\psi_i^{*(m)} = \hat{m}_i(\sobs) + (\psi_i^{(m)} - \hat{m}_i(s^{(m)})) \ \frac{\hat{\sigma}_i(\sobs)}{\hat{\sigma}_i(s^{(m)})}, \quad i=1,\dots, p.
\end{equation}




\textcolor{black}{Considering} nonlinear\textcolor{black}{ity} and heteroscedastic\textcolor{black}{ity allows for more flexibility than the linear case}, albeit at a heavier computational cost. Moreover, the use of feed-forward neural networks can be seen as a dimension reduction stage, since the model can be expressed as a function of the different hidden units whose dimension is in generally much smaller than that of the summary statistics. \textcolor{black}{On the other hand, since random forests are more robust to the presence of irrelevant predictors, they can naturally handle high dimensional statistics.}

Other approaches have been suggested, \textcolor{black}{for example best subset selection methods, projection techniques \citep{Fearn12} or partial least squares approaches, but they were difficult to implement if not unfeasible in our context due to the high dimensionality of our summary statistics.}


%
\textcolor{black}{Linear and nonlinear regression adjustments are available in the \texttt{R} package \texttt{abc}, and the adaptive two-stage nonlinear approach can be implemented using the \texttt{abc} package and the \texttt{svm} function from the \texttt{e1071} package. Package \texttt{ranger} can be used for random forests regression.}

\subsubsection{Second point of view: approximation of unidimensional quantities from the ABC posterior of the summary statistics}\label{sec:ml}

Approaches listed in the previous section make use of regression techniques to produce ABC posterior samples. In this section, \textcolor{black}{we adopt another point of view and} explore a set of methods focusing on the approximation of one-dimensional quantities of interest from the ABC posterior. It can include for example posterior mean or posterior quantiles. The main idea is to build a nonlinear regression model using techniques which can handle a large number of explanatory variables. In \cite{Rayn19}, the authors suggested the \textcolor{black}{use of quantile regression via} of random forests, \textcolor{black}{for their ability to handle high dimensional data as mentioned previously.} 

We propose here a second approach based on gradient boosting. The objective of boosting methods is to build a strong learner from a set of weaker learners, in a sequential fashion. A first regression model is built between the posterior quantity of interest (e.g. the mean or median) and the summary statistics. A second regression model is then built, which focus on the points which were incorrectly predicted by the first regression model. More precisely, weights are assigned to each point, proportionally to the associated quality of prediction: the smaller the prediction error for a given point, the smaller its weight. The process is iterated several times. \textcolor{black}{The prediction error is defined through a loss function measuring the discrepancy between observations and predictions.}
\textcolor{black}{Boosting has already been applied in the context of ABC in \cite{Aes12}, where the authors used different boosting algorithms to learn the nonlinear regression function relying each model parameter and the set of summary statistics. Their approach is similar to those mentioned in section \ref{sec:dimred}. In this paper, we use gradient boosting in order to directly infer the median, mean and selected quantiles of the posterior distribution.}
Here, we combined gradient boosting with quantile regression to estimate the posterior median as well as the posterior quantiles of order 2.5\% and 97.5\%. We also estimated the posterior mean using $L_1$ and $L_2$ loss functions, i.e. minimizing respectively the absolute error and the squared error between predictions and observations in the regression model. \textcolor{black}{The $L1$ loss is known to be less sensitive to outliers \citep{Hastie}.}

ABC via RF is available in the \texttt{R} package \texttt{abcrf}, and ABC via gradient boosting can be implemented using the \texttt{R} package \texttt{gbm}.

\subsection{Simulation study}
The different calibration methods were first compared on simulated data according to the following scheme: (i) $M=100 \ 000$ parameter values were sampled from the prior distributions, and $M$ simulated datasets were generated from these parameter values, (ii) 100 of these datasets were randomly chosen to act as reference datasets, and (iii) ABC posterior samples and quantiles were estimated for each of these 100 reference datasets using the remaining 999 900 datasets, using each method listed in Table \ref{tab:summet}. We compared two different values for the threshold $q_\varepsilon$, using $\varepsilon = 2.5\%$ (resp. 5\%) of the samples, i.e. a final sample size of $N_\varepsilon = 2500$  (resp. 5000).

To compare performances, we computed the relative absolute error (RAE) between the true parameter value and the posterior median respectively. We also computed the empirical coverage of the 95\% CI computed using each approach, defined as the proportion of time the \textit{true} parameter value felt inside the 95\% CI derived from the ABC posterior distribution computed on the associated simulated dataset.
To account for numerical and computational issues encountered with some approaches, we computed the proportion of cases for which each approach failed (due for example to non-convergence or ill-defined estimates).


\subsection{\textcolor{black}{Application to real data}}
\textcolor{black}{In order to assess the performance of each method on real data, we computed the 95\% CI for all parameters, to identify parameters for which the posterior distribution (and hence the 95\% CI) is significantly different from the prior distribution. Then, we also conducted posterior predictive checks, but only on the best method(s) to reduce the computation time. }

\textcolor{black}{There are two cases for the posterior predictive check: either the selected method produces samples from the ABC posterior distribution (for example for rejection methods, or methods based on local regression), or it produces estimates of key quantities from the ABC posterior distribution (for example for quantile regression via random forests or gradient boosting). In the former case, those samples can be used directly to produce predictions from the generative model that will be compared to the observed data. In the latter case, an approximation of the ABC posterior can be computed based on the three quantiles of order 2.5\%, 50\% and 97.5\% using for example the generalized lognormal distribution \citep{Mye19}.
A random variable is said to be distributed as a GLN if it can be written as $c X + d$, where $X$ has a normal or lognormal distribution. It thus encompasses normal and lognormal distributions, with a high flexibility to handle skewness. It can be defined using the three aforementioned quantiles \citep{qpd,perepolkin2021hybrid}. Samples from the approximate ABC posterior were generated, and used to build predictions. }

\section{Results}\label{sec:results}

\subsection{Results of the simulation study}

\begin{figure}
\centering
\includegraphics[scale=0.82]{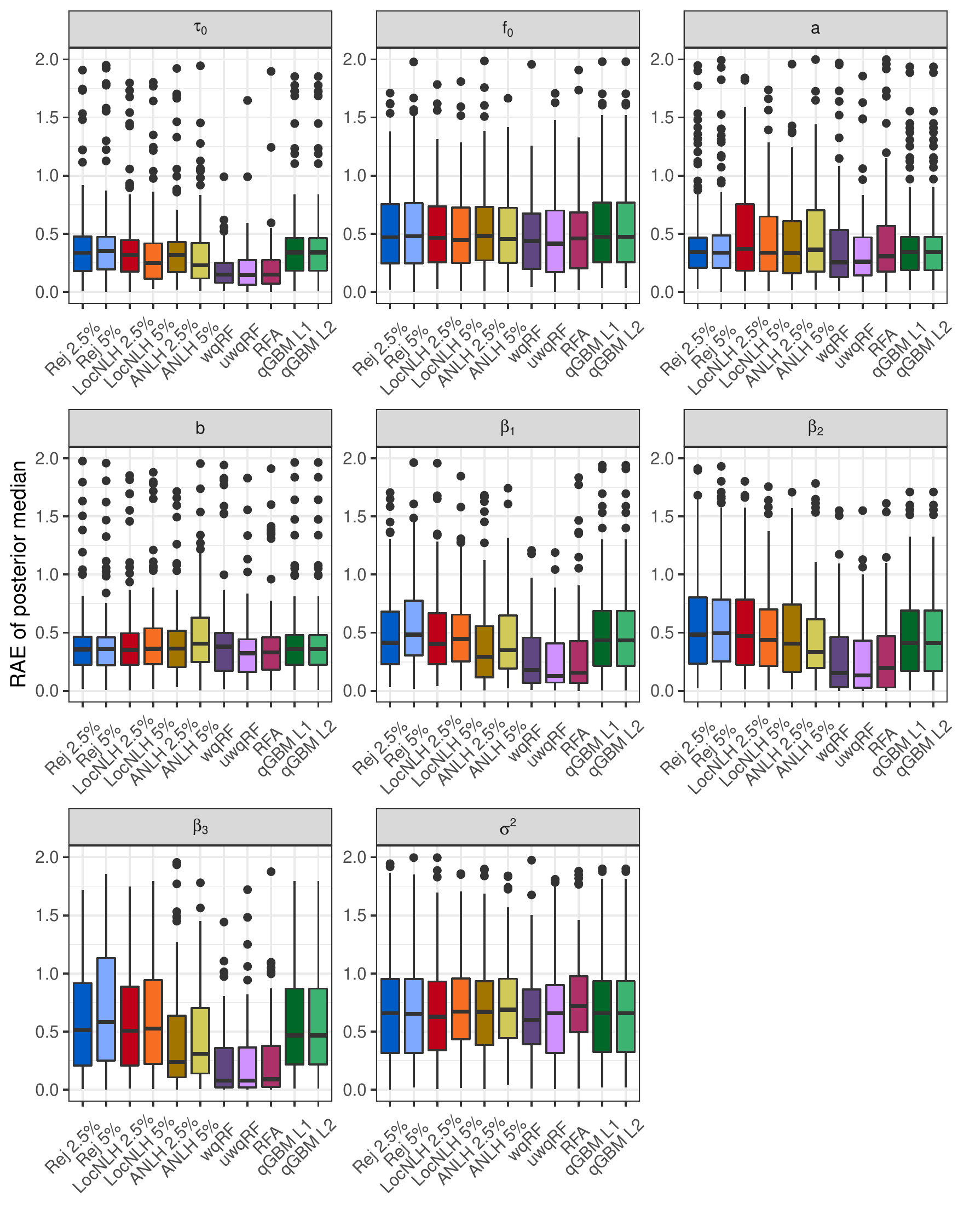}
\caption{Relative absolute error (RAE) of the posterior median on simulated data. The y-axes were truncated to 2 to make the plots clearer by removing some extreme points. Local linear approaches were omitted. \textcolor{black}{Abbreviations (see Table \ref{tab:summet}): `Rej`: rejection ABC, `LocNHL`: local linear regression with heteroscedastic error, `ANHL`: adaptive non linear local regression with heteroscedastic error, `wqRF` (resp. `uwqRF`): weighted (resp. unweighted) quantile regression via random forests, `RFA`: nonlinear regression using random forests, and `qGBM L1` (resp. `qGBM L2'): quantile regression using gradient boosting and $L_1$ (resp. $L_2$) loss. The `2.5\%` and `5\%` terms refer to the threshold used in the method and corresponds to the proportion of simulated parameters which are kept for the analysis.}}
\label{fig:rae_median}
\end{figure}

Figure \ref{fig:rae_median} provides RAE of the ABC posterior median obtained with each method (the corresponding figure for ABC posterior mean is Figure 1 
in the Supplementary material). 
Overall, we obtained RAE ranging from 0 to 2 (excluding too extreme values on the graphs), with most of the values lying between 0.25 and 0.75. Random forests \textcolor{black}{approaches (either producing adjusted samples or approximating univariate posterior quantities)} were associated with the smallest RAE values, for all the parameters considered. The adaptive nonlinear heteroscedastic models performed generally better than non adaptive ones, but no clear difference can be found between the two thresholds $2.5\%$ and $5\%$. This is consistent with previous results on the effect of the threshold in regression-based approaches \citep{Beau02}. Local linear approaches lead to ABC posterior distributions with very large ranges of variations, rarely respecting the support constraints given by the prior distributions. For example, large negative values were obtained for parameters which are positive by definition. For a large proportion of datasets (depending on the method and parameter, between 15\% and 25\%), these approaches actually failed numerically and outputed infinite values (see Table \ref{tab:covCI} and Table 1 
in the Supplementary material).  As a result of these erratic estimations, the RAE associated to these two approaches was also highly variable. For the sake of clarity, these results were thus excluded from graphical representations.

Empirical coverages are given in Table \ref{tab:covCI}. \textcolor{black}{Generally, the empirical coverages are closer to the nominal level when the threshold $\varepsilon$ is smaller. Overall, random forest approaches performed best in terms of both empirical coverage and numerical stability, while local regression methods performed worst, with too low empirical coverages and a high failure rate. } For each parameter, we computed the rank of each method, assigning rank 1 to the one having the closest empirical coverage to the theoretical value of 0.95. The average rank across parameters is given in the last column of Table \ref{tab:covCI}. Using this criterion, the best approaches are random forests and gradient boosting.  \textcolor{black}{Their better performances might be partly explained by the fact that quantile regression approaches naturally focus on the estimation of bounds for the credible interval and therefore might perform better for that task. In contrast, approaches relying on the extraction of ABC posterior quantiles from a set of ABC posterior samples might be less accurate. However, it is worth noting that RFA, which also produces adjusted samples from the ABC posterior performs better than other regression-based methods, with performances similar to those obtained with quantile regression approaches. The simple ABC rejection algorithm performs better than local regression approaches in terms of empirical coverage. It is noteworthy however that due to the sampling scheme of the simulation study, since the ``true'' parameter values behind each dataset were randomly sampled from the prior, a method producing ABC posterior distributions that resemble the prior distribution might have advantages. Indeed, they would naturally lead to 95\% CI containing the true value.} 

\begin{table}[ht]
\centering
\caption{Empirical coverages based on the 95\%CI estimated using each approaches (proportion of datasets for which the true value used to generate the simulated data felt inside the estimated 95\% CI), average rank and average proportion of model failure (proportion of datasets for which each approach failed, averaged over the parameters).  \textcolor{black}{Abbreviations (see also Table \ref{tab:summet}): `Rej`: rejection ABC, `LocNHL`: local linear regression with heteroscedastic error, `ANHL`: adaptive non linear local regression with heteroscedastic error, `wqRF` (resp. `uwqRF`): weighted (resp. unweighted) quantile regression via random forests, `RFA`: nonlinear regression using random forests, and `qGBM L1` (resp. `qGBM L2'): quantile regression using gradient boosting and $L_1$ (resp. $L_2$) loss.}}
\small{
\begin{tabular}{lcrrrrrrrr>{\centering\arraybackslash}m{1.5cm}>{\centering\arraybackslash}m{1cm}}
  \hline
Method & $\varepsilon$ & $\tau_0$ & $f_0$ & $a$ & $b$ & $\beta_1$ & $\beta_2$ & $\beta_3$ & $\sigma^2$ & Average rank & Average failure\\ 
  \hline
Rej & 2.5\% & 0.918 & 0.969 & 0.938 & 0.928 & 0.938 & 0.959 & 0.979 & 0.928 & 3.125 & 0\\ 
  Rej & 5\% & 0.907 & 0.979 & 0.938 & 0.928 & 0.928 & 0.969 & 0.979 & 0.938 & 4.875 & 0\\ 
  LocLH & 2.5\% & 0.830 & 0.772 & 0.778 & 0.763 & 0.806 & 0.813 & 0.767 & 0.815 & 10.5 & 26\\ 
  LocLH & 5\% & 0.732 & 0.760 & 0.760 & 0.821 & 0.729 & 0.792 & 0.760 & 0.779 & 12.5 & 19 \\ 
  LocNLH & 2.5\% & 0.897 & 0.948 & 0.825 & 0.814 & 0.918 & 0.897 & 0.969 & 0.907 & 5.75 & 0\\ 
  LocNLH & 5\% & 0.887 & 0.948 & 0.804 & 0.784 & 0.856 & 0.938 & 0.948 & 0.887 & 7.375 & 0 \\ 
  ANLH & 2.5\% & 0.732 & 0.845 & 0.825 & 0.773 & 0.794 & 0.753 & 0.887 & 0.845 & 10.375 & 0 \\ 
  ANLH & 5\% & 0.825 & 0.845 & 0.742 & 0.722 & 0.763 & 0.804 & 0.856 & 0.804 & 12.125 & 0 \\ 
  wqRF & 5\% & 0.940 & 0.960 & 0.920 & 0.920 & 0.920 & 0.980 & 0.980 & 0.900 & 5.5 & 0 \\ 
  uwqRF & - & 0.960 & 0.970 & 0.940 & 0.950 & 0.950 & 0.980 & 0.990 & 0.940 & 3.25 & 0 \\ 
  RFA & 5\% & 0.970 & 0.950 & 0.960 & 0.940 & 0.990 & 0.970 & 0.930 & 0.970 & 4.375 & 0\\ 
  qGBM L1 & 5\% & 0.907 & 0.979 & 0.938 & 0.928 & 0.959 & 0.969 & 0.979 & 0.938 & 4.125 & 2.37 \\ 
  qGBM L2 & 5\%  & 0.907 & 0.979 & 0.938 & 0.928 & 0.959 & 0.969 & 0.979 & 0.938 & 4.125 & 5 \\ 
   \hline
\end{tabular}
}
\label{tab:covCI}
\end{table}

\begin{figure}[h]
	\centering
	\includegraphics[scale=0.6]{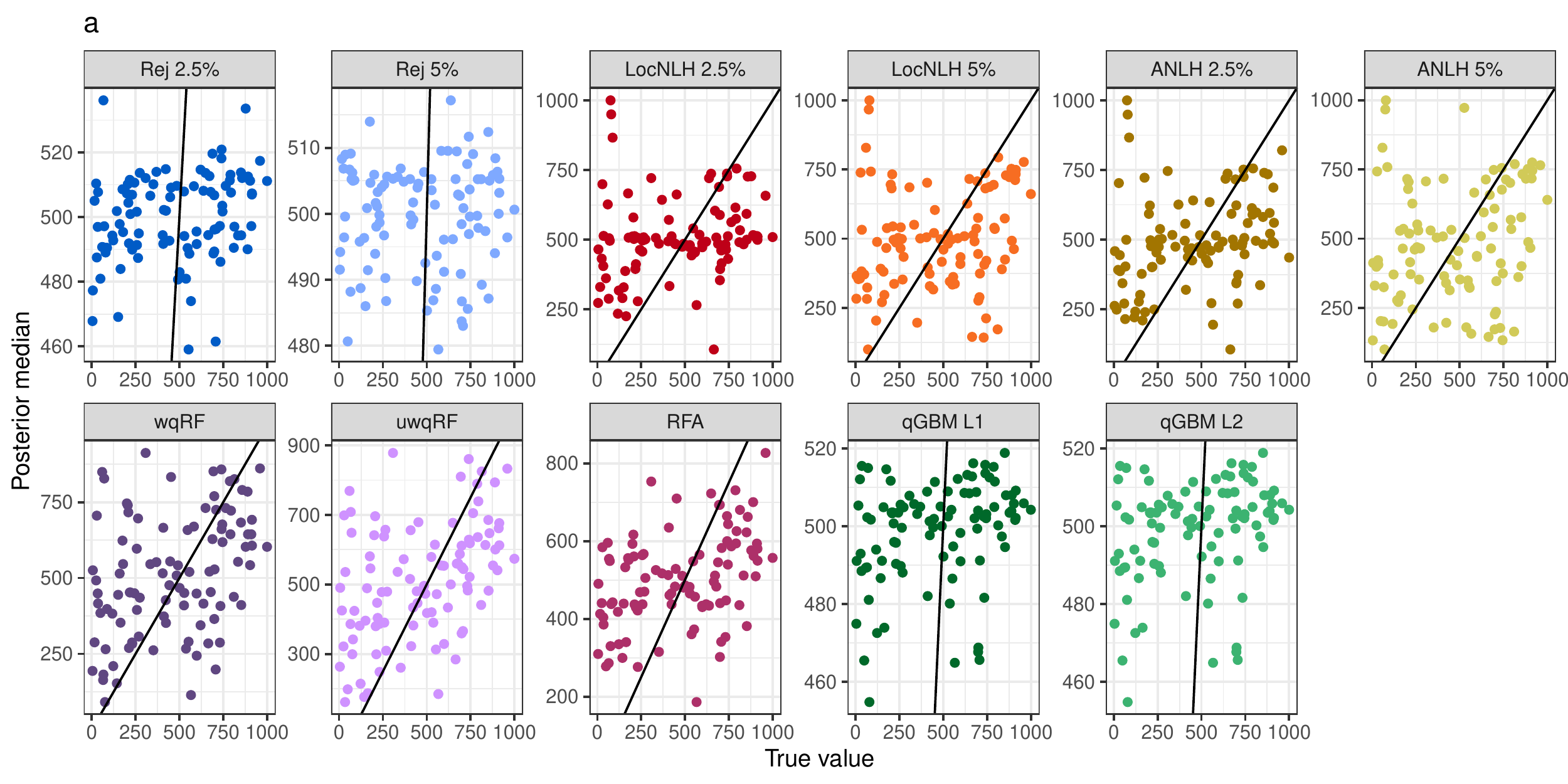}
	\includegraphics[scale=0.6]{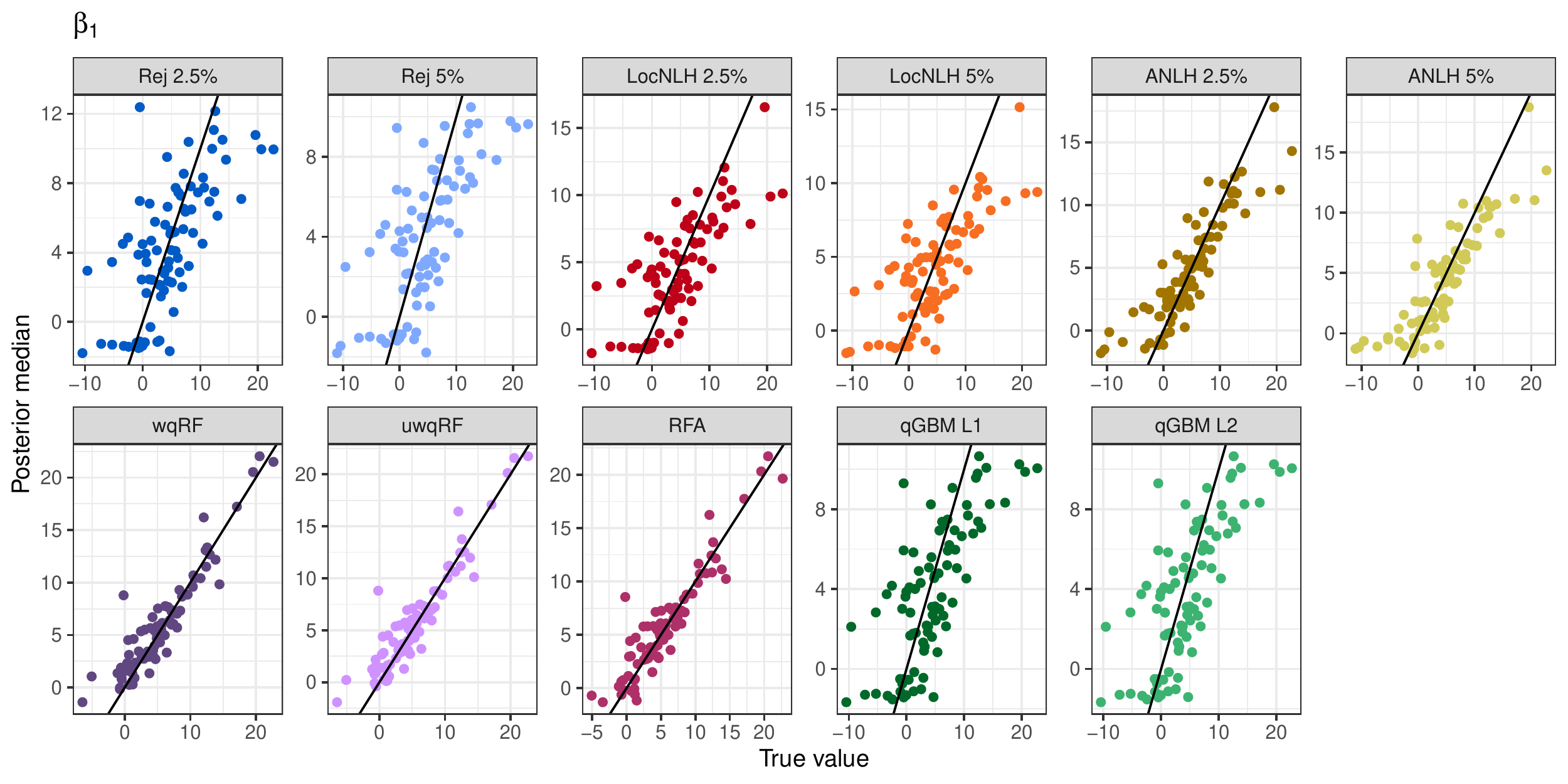}
	\caption{ABC posterior median as a function of the true parameter value for parameters $a$ and $\beta_1$. Abbreviations (see also Table \ref{tab:summet}): `Rej`: rejection ABC, `LocNHL`: local linear regression with heteroscedastic error, `ANHL`: adaptive non linear local regression with heteroscedastic error, `wqRF` (resp. `uwqRF`): weighted (resp. unweighted) quantile regression via random forests, `RFA`: nonlinear regression using random forests, and `qGBM L1` (resp. `qGBM L2'): quantile regression using gradient boosting and $L_1$ (resp. $L_2$) loss. The `2.5\%` and `5\%` terms refer to the threshold used in the method and corresponds to the proportion of simulated parameters which are kept for the analysis.}	
	\label{fig:med_true_param}
\end{figure}

For each method, we also compared the ABC posterior median with the true value of the parameter. \textcolor{black}{Overall, the performances of each approach vary across parameters, with random forests providing the best results. Posterior medians computed with ABC rejection algorithms poorly reflected the true underlying parameter values. This might be due to the fact that rejection methods produced posterior medians which were mostly located around the prior median.} Observation parameters were better estimated than model parameters. As an example, Figure \ref{fig:med_true_param} gives the estimated versus true parameter value for $\beta_1$ \textcolor{black}{for which the methods provided good results} and for $a$ \textcolor{black}{for which the methods performed poorly}. A complete plot for all the parameters is given in the Supplementary material (Figure 2
). Methods based on simple rejection do not capture the full range of variability of parameter $a$, and provide a posterior mean which is too close from the prior mean. Local nonlinear heteroscedastic approaches are able to cover a larger range of variability, but are still too focused on the prior mean. The two-stage adaptive approach allows for better results, which are comparable to those obtained with random forests. Gradient boosting suffers from the same flaws as rejection approaches for this parameter. In contrast, parameter $\beta_1$ is better estimated, independently of the method used. Random forests performs remarkably well for this parameter \textcolor{black}{and are more efficient to estimate parameters that are located far from the prior median.}

\subsection{Results using field data}

In this section, we detail the results obtained on field data. 
Figure \ref{fig:ci_real} provides the median and 95\% CI for each method and parameter (see also Table 2 
 in the Supplementary material). Contrary to what was observed in the simulation study, very similar results were obtained for both rejection approaches, suggesting a small effect of the threshold $\varepsilon$. For local nonlinear approaches, there was an effect of the choice of $\varepsilon$ on the results, especially for parameters $\tau_0, a,b$ and $\beta_2$. Results were more consistent between adaptive and non adaptive methods for the same threshold value, than within the same method but for different threshold values. Results based on \textcolor{black}{quantile regression via} random forests were consistent whether weighted or unweighted samples were used, and were consistent with results obtained with nonlinear approaches for $\varepsilon=5\%$.  Results obtained using gradient boosting were similar, whatever the loss function that was used. Methods based on random forests exhibit large variance in \textcolor{black}{some} the ABC posterior distributions.

\begin{figure}[h]
\centering
\includegraphics[scale=0.8]{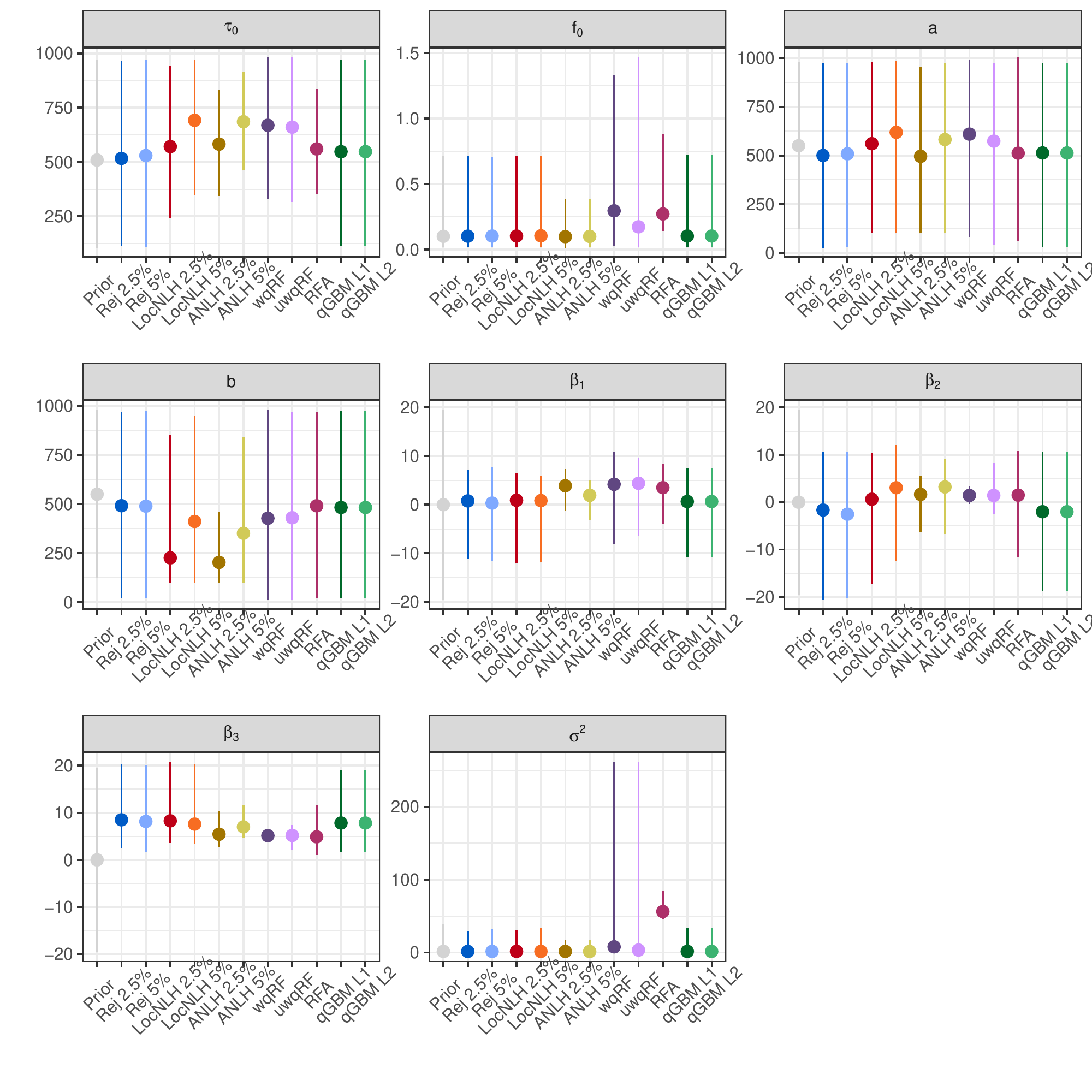}
\caption{95\% ABC credible interval of each method on the field data. Abbreviations (see also Table \ref{tab:summet}): `Rej`: rejection ABC, `LocNHL`: local linear regression with heteroscedastic error, `ANHL`: adaptive non linear local regression with heteroscedastic error, `wqRF` (resp. `uwqRF`): weighted (resp. unweighted) quantile regression via random forests, `RFA`: nonlinear regression using random forests, and `qGBM L1` (resp. `qGBM L2'): quantile regression using gradient boosting and $L_1$ (resp. $L_2$) loss. The `2.5\%` and `5\%` terms refer to the threshold used in the method and corresponds to the proportion of simulated parameters which are kept for the analysis.}
\label{fig:ci_real}
\end{figure}

Overall, results differed according to the type of parameters considered. On the one hand, for model parameters i.e. $\theta= (\tau_0,f_0,a,b)$, rejection methods and gradient boosting approaches yielded credible intervals which were very similar to those derived from the prior distributions, which was not the case for local nonlinear approaches and RF methods. Parameter $a$  was the most difficult to estimate, and little additional information was conveyed by the ABC posterior distributions or by credible intervals, compared to the information provided by the prior distribution. On the other hand, for observation parameters i.e. $\omega = (\beta_1,\beta_2,\beta_3,\sigma^2)$, all methods produced 95\% CI which significantly differed from the 95\% interval provided by the prior distributions. 
The 95\% CIs revealed that there is a high uncertainty for some parameters, especially for parameters $a$ and $b$, which were already identified as difficult to estimate in the simulation study. The residual variance $\sigma^2$ was also estimated with a large credible interval. This was already the case in the simulation study, where the ABC posterior distributions for this parameter were highly asymmetric with heavy right tails (a shape also provided via the prior). For this parameter, the different approaches gave similar ABC posterior medians, except random forests methods (see Table 2 
in the Supplementary material).

\begin{table}[ht]
\centering
\small{
\caption{Posterior median and 95\% CI for each parameter using the quantile regression approach and the nonlinear regression approach based on random forests, and the adaptive nonlinear local approach (with $\varepsilon=2.5\%$), on the field data.}
\begin{tabular}{>{\centering\arraybackslash}m{1.5cm}>{\centering\arraybackslash}m{4.25cm}>{\centering\arraybackslash}m{4.25cm}>{\centering\arraybackslash}m{5cm}}
  \hline
Parameter & Quantile regression using unweighted random forests & Nonlinear regression using random forest & Adaptive nonlinear heteroscedastic regression, $\varepsilon=2.5\%$ \\
\hline
$\tau_0$ & 660 [317 ; 982] & 560 [352 ; 837] & 583 [345 ; 832]\\
$f_0$  & 0.17 [0.02 ; 1.47] &  0.27 [0.14 ; 0.88] & 0.097 [0.013 ; 0.39] \\
$a$ & 574 [39 ; 975] & 512 [61 ; 1005] & 496 [100 ; 955]\\
$b$ & 430 [12 ; 966] & 491 [20 ; 970] & 203 [100 ; 460]\\
$\beta_1$ & 4.40 [-6.46 ; 9.55] & 3.50 [-3.85 ; 8.38] & 3.89 [-1.35 ; 7.33]  \\
$\beta_2$ & 1.42 [-2.41 ; 8.28] & 1.48 [-11.52 ; 10.85] & 1.65 [-6.34 ; 5.68] \\
$\beta_3$ & 5.19 [2.12 ; 7.39] & 4.90 [1.07 ; 11.63] & 5.44 [2.68 ; 10.4] \\
$\sigma^2$ & 3.28 [0.31 ; 261] & 56 [45 ; 85] & 1.44 [0.29 ; 17]\\ 
  \hline
\end{tabular}
\label{tab:res_uqrf}
}
\end{table}

Based on the simulation study and previous remarks, and to lighten the presentation, we focus on three approaches for generating predictions: the adaptive local nonlinear method and random forests approaches. \textcolor{black}{For these approaches, the posterior medians and the 95\% CI are reported in Table \ref{tab:res_uqrf}.} 
The other approaches either provided ABC posterior distributions that were too close from the prior distributions (e.g. for parameters $\tau_0$, $f_0$, $a$ and $b$), or had too large credible intervals (e.g. for parameters $\beta_1$, $\beta_2$, $\beta_3$ and $\sigma^2$).
The latter is to some extent also true for random forests methods (for parameters $f_0$ and $\sigma^2$ for example), but this is balanced by obtaining smaller credible intervals for other parameters, and by the promising results in the simulation study. For the ANLH approach, we chose the threshold $\varepsilon=2.5\%$ due to the smaller RAE found on the simulated data, and for the random forest approach we chose \textcolor{black}{the approach producing adjusted samples and the quantile regression via unweighted random forests.}

The 95\% credible intervals obtained for $\beta_1$ and $\beta_2$ included 0 (see Table \ref{tab:res_uqrf}), which means that we could not reject the hypothesis that there is no effect of periods 1 and 2 on the visitation rate intensity. On the other hand, results suggested that the intensity increases in the third period, which might result from an increase in the abundance of workers at the end of the season\textcolor{black}{, as well as foraging by dispersing drones and new queens.}

\begin{figure}[t]
\centering
\subfloat[Variable plot]{\includegraphics[scale=0.7]{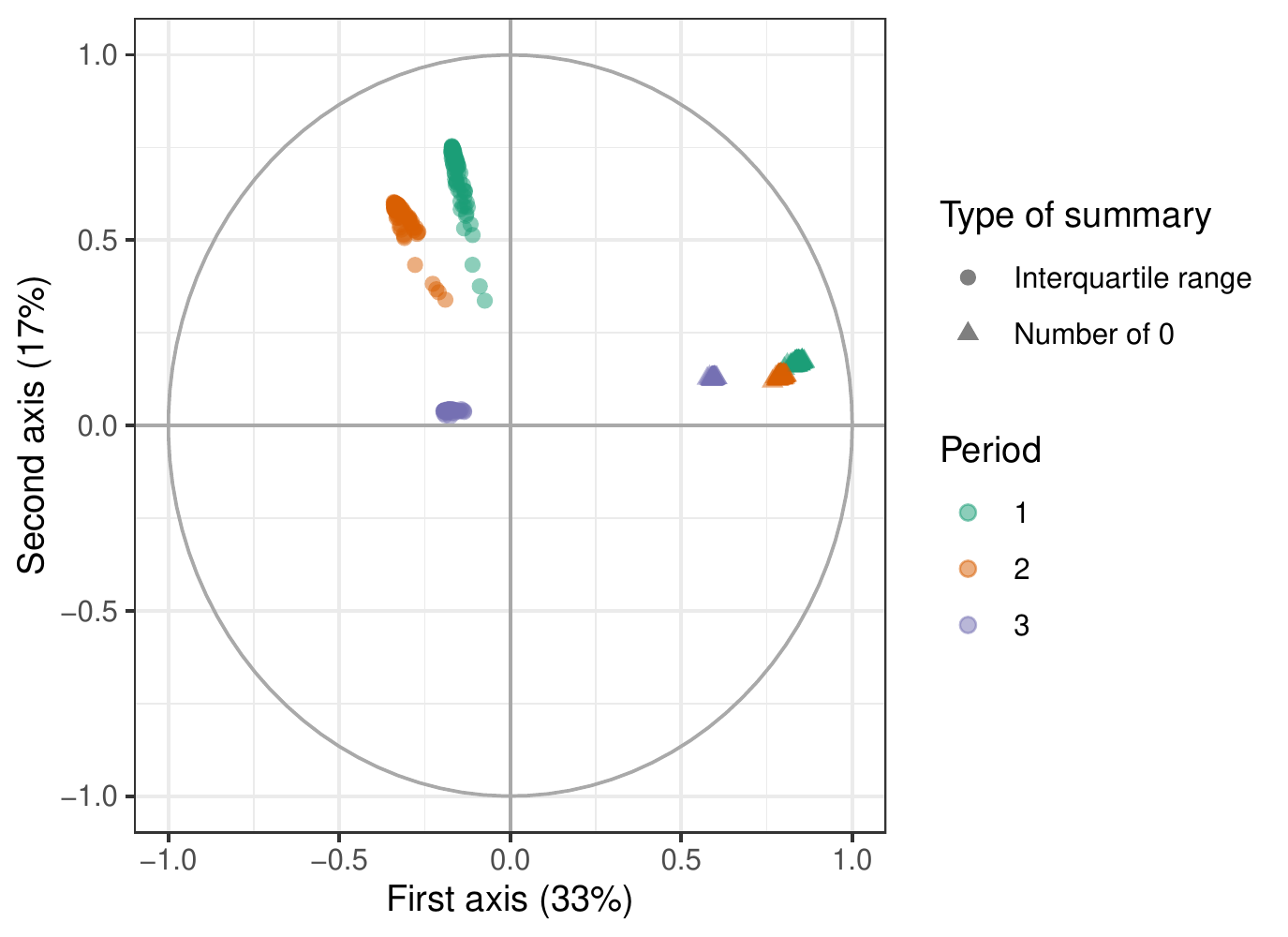}}
\subfloat[First PCA plane]{\includegraphics[scale=0.7]{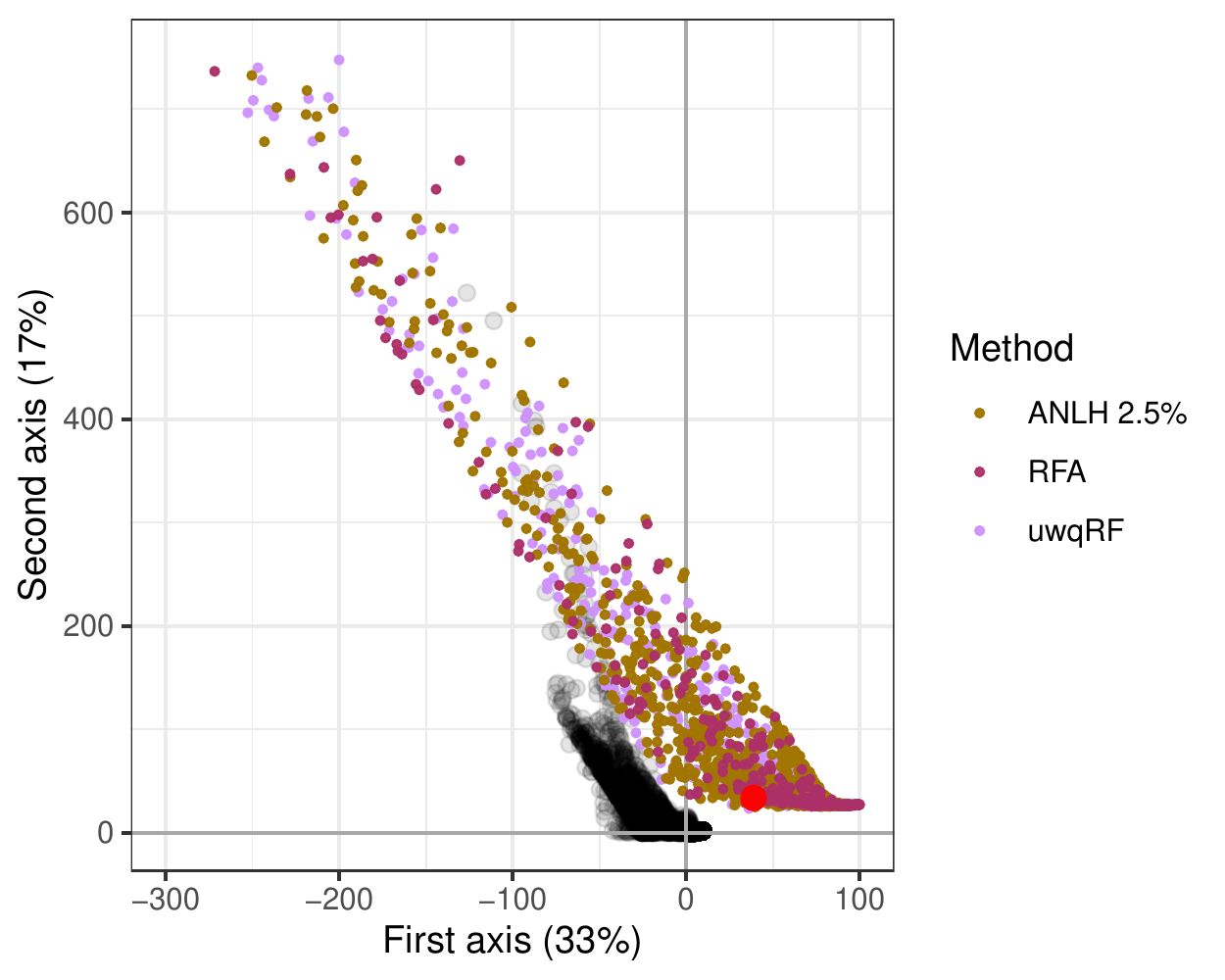}}
\caption{Distribution of the summary statistics from the ABC table (in black dots) in the first two planes of a PCA. The red dot corresponds to the projection of the observed summary statistics in the same planes, and the other colored dots correspond to the distribution, in the same planes, of the predicted summary statistics obtained using quantile regression via unweighted random forests (uwqRF), adaptive nonlinear local regression (ANLH) and nonlinear regression via random forest (RFA). Axes were truncated to enhance readability.}
\label{fig:pca_pred}
\end{figure}

\textcolor{black}{To compare predicted and observed values, we performed a principal component analysis (PCA) on the summary statistics of the ABC table. Figure \ref{fig:pca_pred} represents the distribution of the summary statistics from the ABC table and from the predicted distributions obtained with uwqRF, ANLH and RFA, in the first plane of the PCA. Contrary to the prior distribution of the summary statistics, the predicted distributions are located around the observed summary statistics, but with very high tails. This can also be seen in Figure \ref{fig:hist_pca} for the first axis (and in the Supplementary material for the second and third axes). }
Overall, the different approaches produced over-dispersed predictions compared to the observations. \textcolor{black}{This is particularly the case along a direction corresponding to the inter-quartile ranges during the first and second periods. Indeed, some of the predictions were unrealistically high, leading to extreme values for the interquartile range.}
 The range of variation was greater with methods based on random forests, and in general the variance of the predictions was higher than with ANLH.

\begin{figure}
\centering
\includegraphics[scale=0.8]{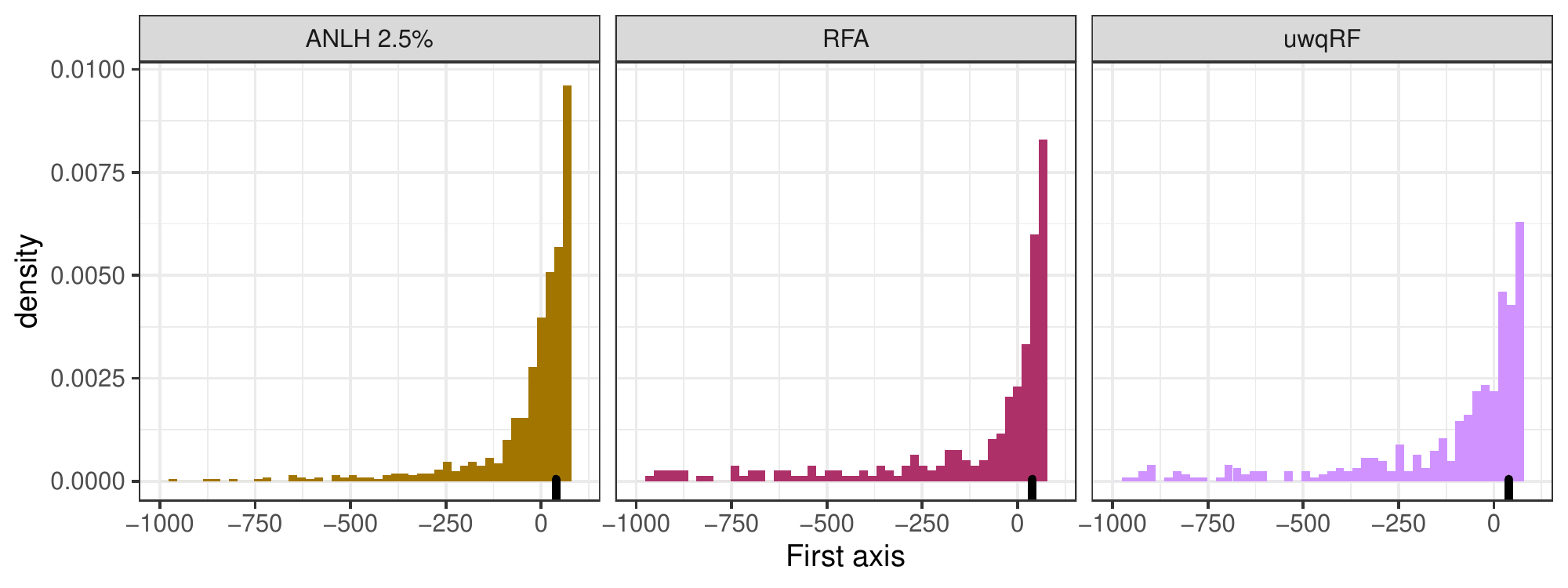}
\caption{Histogram of the summary statistics along the first axis of the PCA (left panel: adaptive nonlinear local regression (ANLH), middle panel: nonlinear regression via random forests (RFA) and right panel: quantile regression via unweighted random forests. The black segments correspond to the location of the observed summary statistics on the first PCA axis.}
\label{fig:hist_pca}
\end{figure}

\begin{figure}[h]
\centering
\includegraphics[scale=0.75]{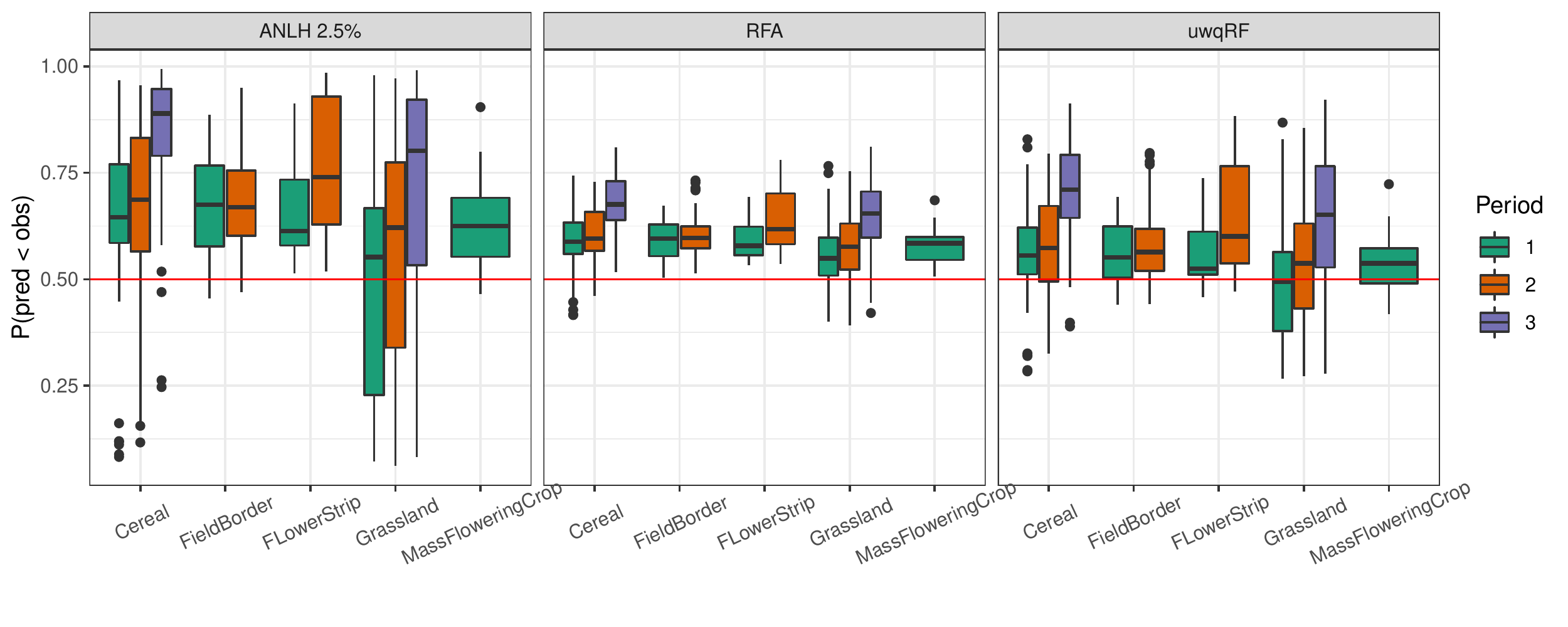}
\caption{Probability that the predicted data (left panel: obtained using adaptive nonlinear regression, right panel: obtained using nonlinear regression via random forests) fall below the observed data, for each landuse type and period. Red horizontal lines correspond to the 2.5\% and 97.5\% levels.}
\label{fig:quantObs}
\end{figure}

We then for each data point computed the probability for the predicted data to be smaller than the observed data. This probability can be seen as a Bayesian $p$-value and it is expected that in the absence of systematic under- or over-estimation, its distribution should be uniform over $[0,1]$. Better results were obtained with uwqRF (see Figure \ref{fig:quantObs}). In general, the predictions tended to underestimate the observations, especially with ANLH. \textcolor{black}{This is due to a high number of 0s in the predictions (this can also be seen in Figure 3 
in the Supplementary material, where the predictions are compared to observations for some randomly selected sites.)}

\begin{figure}
\centering
\subfloat[Nesting (left panel) and floral maps (right panel). Nesting values are either 0 or 1 (absence/presence of a nest)]{\includegraphics[scale=0.5]{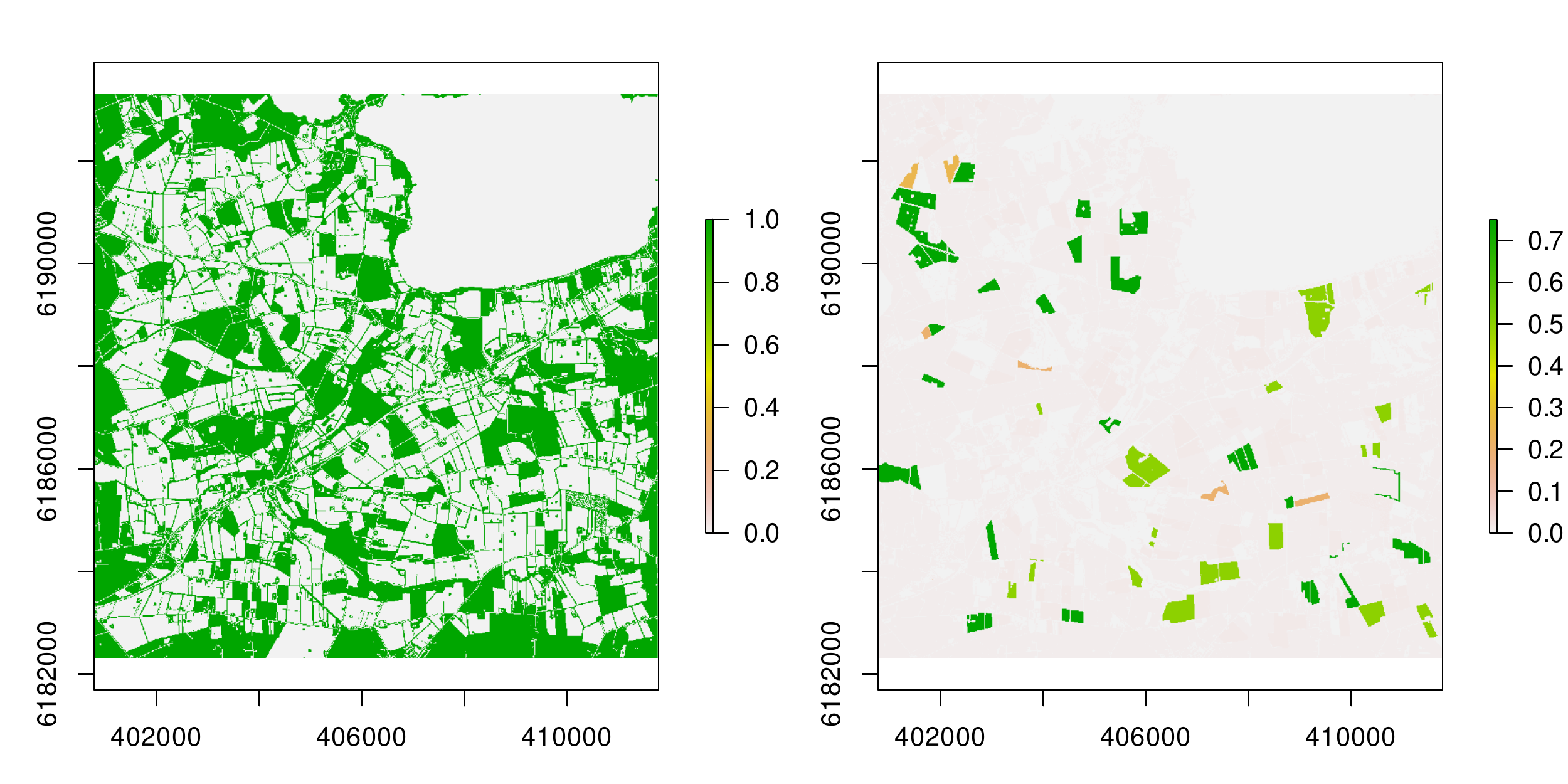}}

\subfloat[Predictions from the model (left panel: uwqRF, middle panel: ANLH, right panel: RFA)]{\includegraphics[scale=0.48]{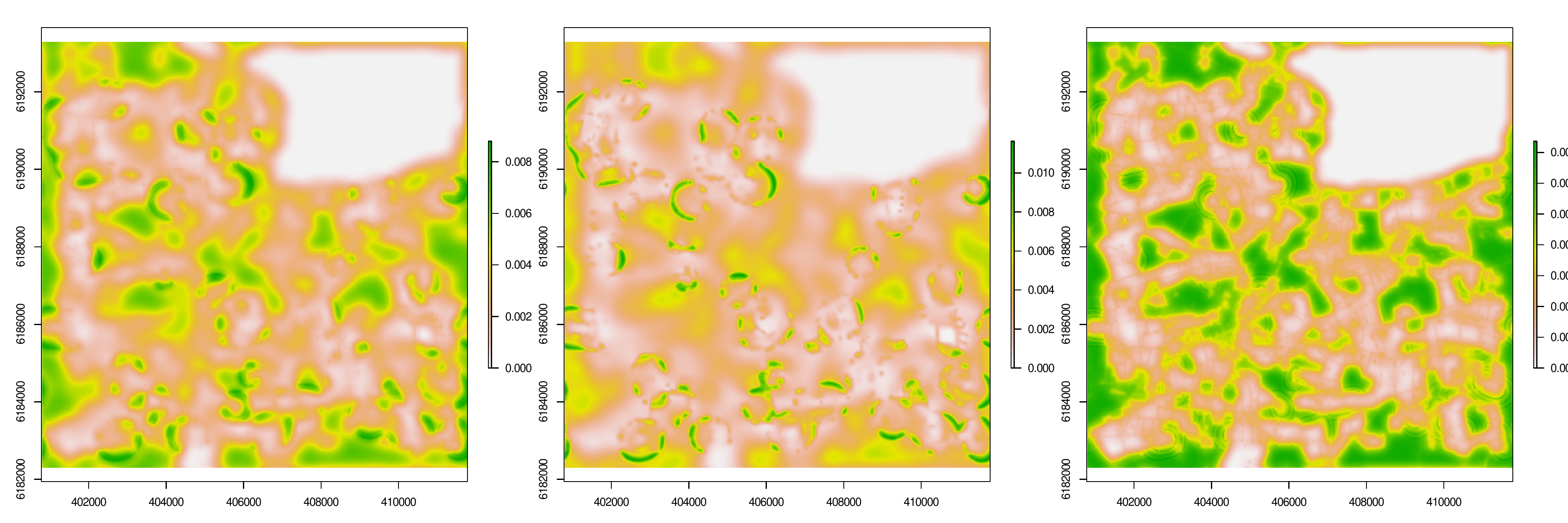}}
\caption{Predicted visitation rate intensity on a landscape (a map of 10km$\times$10km), for a given year and a given period. Input nesting and floral maps are provided in the upper row, while predictions are given in the lower row.}
\label{fig:pred_maps}
\end{figure}

\textcolor{black}{Due to the computation time required to run the model on all the landscapes, it does not seem reasonable to compute predictions of the number of bees based on multiple runs of the model using different parameter values from the posterior distributions. However, posterior medians can be used in a first approach. Figure \ref{fig:pred_maps} gives an example of predictions from the model using calibrated parameter values.}

\section{Discussion and conclusion}


In this paper, we proposed a methodology to perform a Bayesian calibration of parameters from a complex nonlinear ecological model where the likelihood is intractable. An Approximate Bayesian Computation (ABC) approach was used, and several methods and algorithms were compared for the estimation of the ABC posterior distribution. A set of summary statistics was used in place of the original data to reduce the dimension of the problem as well as noise, while also introducing a bias which decreases when the summary statistics convey enough information about the data.  \textcolor{black}{We showed that ABC was able to provide valuable information about the posterior distribution of the parameters, confirming previous promising results on the use of ABC in the context of ecological models \citep{van2015calibration}}.

\textcolor{black}{Some parameters were easier to estimate than others. More specifically, we found that model parameters, i.e. those used in the CPF model, were more difficult to estimate, especially with local approaches. This might be due to the fact that the summary statistics do not convey enough information for these parameters, or that the original dataset in itself is not informative enough leading to practical unidentifiability. It might also be due to a small sensitivity of these parameters on the model outputs.  On the other hand, observation parameters were easier to identify, except for local approaches and simple rejection with which the posterior median seems to be localized around two modes, illustrating the limits of local approaches.}

\textcolor{black}{As far as the CPF model is concerned, results are encouraging, since the posterior distribution of most parameters was narrower than the associated prior distribution, which means that the data conveyed enough information about these parameters. Our aim was not to test or validate the CPF model \textit{per se}, but rather to propose a methodology to calibrate this type of ecological model. A recent study \citep{Nicholson19} compared the performances of two pollination models, namely the CPF model and Lonsdorf's diffusion model, for fixed sets of parameter values, and showed that the CPF model better reflects the spatial dynamic of foraging bees. If the variability of the floral and nesting values needed as inputs for the model were taken into account in the aforementioned study, they compared the models based on the predicted intensity, while we added an observation process to move from intensity to visitation rates. Moreover, we considered the parameters as unknown and not as fixed to reference values. 
Wide uncertainty in the calibrated parameters and poor predictive performance of the calibrated model could result from shortcomings of the CPF model, to a noisy observation process, or both.
 Our study can still be seen as a first step towards the calibration of spatially-explicit pollination models, even though more efforts are needed. For example, equation \eqref{eq:traveldist} is a simplified version of the actual dynamic response of the nest-specific distance to the surrounding landscape that is used in the original CPF model, which is easier to parameterize, but the two versions have not been compared and tested on actual data (which ideally would require calibration). The two parameters involved in the nest-specific dynamic, $a$ and $b$ were also poorly estimated, suggesting some kind of practical unidentifiability.}


\textcolor{black}{
As far as the statistical methodology is concerned, three competing methods emerged from our simulation study: quantile regression via random forests (uwqRF), and regression adjustment of posterior samples via adaptive nonlinear local regression (ANLH) or random forests (RFA), even though they tended to provide over-dispersed predictions. We also observed several very large predicted values, which means that the variability of the predictions was too high. This might be explained by an overestimation of the observation noise variance $\sigma^2$. ABC posterior distributions differed with each approach, even though posterior medians obtained with each method felt within the 95\% CI obtained with the other methods (i.e. results were still consistent). The fact that they produced underestimated predictions might be due to the fact that they predicted 0 bees more often than what was actually observed. This, in turn, could be explained by a too small intensity for the Poisson distribution, which also influence the variance of the predictions.}

\textcolor{black}{On the one hand, unweighted quantile regression via RF is easy to implement, and there is no threshold parameter to tune, but on the other hand it provides only one-dimensional quantities from the ABC posterior, and must be run once for each of these quantities. 
If one is interested in making predictions from the model, it is also necessary to derive a probability distribution from these quantiles, adding an extra level of uncertainty. Other works have been done in the direction of posterior density estimation (see for example \cite{Izbicki19} for an approach based on non-parametric conditional distribution estimation in the context of costly simulations). 
However, if one is only interested in parameter estimation, RF might be more accurate than ANLH, as indicated by the simulation study. Moreover, this method can naturally handle high dimensional summary statistics, so that it might be used when relevant statistics are difficult to identify. On the other hand, the implementation of ANLH approaches is more involved since there is no specific \texttt{R} package, and since they require an additional step for the estimation of the density support. In addition, they rely on the choice of a threshold parameter $\varepsilon$, which was not fully discussed here, but which may have an influence on the results. The advantage of these approaches is that they provide ABC posterior samples, so that predictive distributions are easy to compute. On the field data, they also provided smaller credible intervals for parameters for which RF had some difficulties (e.g. $\sigma^2$).
A possible alternative to these two approaches is RFA, which produces adjusted samples from the ABC posterior. Similarly to uwqRF, it is easy to implement and it naturally handles high dimensional summary statistics ; similarly to regression adjustment approaches, it provides ABC posterior samples. It still relies on a threshold $\varepsilon$ that should be tuned.}


ABC approaches also rely on the definition of a set of summary statistics. In this paper, we derived these summary statistics based on their biological meaning\textcolor{black}{, as explained in Section \ref{sec:sumstat}. Since this choice can be crucial, it deserves some attention}. More automated methods have been developed in the literature \citep{Joy08,Weg09,Nunes10,Fearn12}, but they could not be implemented in our case due to their high computational cost. In high-dimensional settings, \cite{Prang18} recently proposed an approach based on rare events methodology and sequential Monte Carlo, which allow to decrease the computation cost.

\section{Future developments}
Apart from the very nature of ABC which produces an approximation to the true posterior, hence introducing an additional layer of uncertainty, many reasons can explain why the predictions do not reflect the observations with perfect accuracy, and several extensions are possible to enhance the results. 

First, even though the summary statistics were chosen with the objective of extracting meaningful information from the data, sufficiency in the statistical sense (i.e. the fact that the summary statistics carry sufficient information to estimate the model's parameters) is almost impossible to reach. Therefore, part of the information originally contained in the data is missing, which may impact the estimation and as a consequence the predictions. This effect could be partly controlled by adding more summary statistics, but a compromise must be made between computational cost and accuracy. Another issue that we did not discuss here is the fact that we used data from two field studies, i.e. with different recording protocols. This might also have an influence on the results.

Then, the model in itself as well as its inputs, have a great impact on the estimation and prediction processes. For example, we run the CPF model for each floral period and each year separately, only adding a period-specific effect in the observation process. It would be interesting to enrich the model and add a temporal dynamic to account for the growth of the population across the season (see for example \cite{Hau17}). \textcolor{black}{Recent approaches bridging together elements from the CPF theory and ideal-free distribution models have proved to be efficient for modelling honey bee foraging \citep{robinson2022optimal}}. Period- and year- specific land use maps were also used as inputs for the model, and floral and nesting values were generated once at the beginning of the study and then considered as fixed during both the estimation and the prediction processes. This allowed us to account for spatial and temporal variability at the landscape scale, but all our results are then conditional on these realized maps. Due to computational constraints, it was not possible to generalize the process and average the results over several realizations of the land use maps. The importance of floral and nesting values for the calibration of complex pollination model has been acknowledge in \cite{Gardner20}. \textcolor{black}{A more accurate description of the nesting and floral input maps would also reduce the uncertainty related to the landscape. For example, nesting values are currently binary values indicating whether the habitat is suitable for nesting or not. The number of bees per nest is thus the same accross the landscape. It would be interesting to add some variability on this input, or to work in controlled environment where the number of bees nesting at given locations is known. At a much higher computational cost, one might also consider floral and nesting values as parameters, adding them in the ABC process.}

\textcolor{black}{As for the CPF model, it would be interesting to test more sophisticated versions of equation \eqref{eq:traveldist}.}
Other formulations of the statistical model can also be compared: e.g. negative binomial distribution instead of a Poisson, as well as other positive distributions in place of the lognormal one in the definition of the likelihood. \textcolor{black}{This would allow to distinguish the effect of the CPF model and of this observation process.}
Model selection approaches can then be used to identify the best model given the field data. This can be done in an ABC framework using similar approaches as those used in this paper \citep{Prangle14,Pudlo16}, even though care must be taken for this type of analysis \citep{Rob11}.

\section{Author contributions}
CB and US conceived the ideas and designed methodology; HGS and MR collected the data; OO and HGS developed the foraging model, YC and US contributed with land use modelling, CB did the model simulations and analyses; CB led, supported by US and HGS, the writing of the manuscript. All authors contributed critically to the drafts and gave final approval for publication.

\section{Acknowledgements}
CB would like to thank the French National Centre for Scientific Research (CNRS) for financial support through a PEPS-JCJC project.
US was supported by the Swedish research council FORMAS through the project ``Scaling up uncertain environmental evidence'' (219-2013-1271) and the strategic research environment Biodiversity and Ecosystem Services in Changing Climate (BECC).
This project acknowledges funding from European Community's Seventh Framework Programme funded project STEP: Status and Trends of European Pollinators (no. 244090) and Liberation (no. 311781) and from the FORMAS project ``SAPES—Multifunctional agriculture: harnessing biodiversity for sustaining agricultural production and ecosystem services''. \textcolor{black}{Romain Carrié and Johan Ekroos have been helpful in sharing data from the COST project.}
\textcolor{black}{The authors would like to acknowledge comments from two anonymous reviewers that help improving the manuscript, in particular by suggesting the use of random forest adjustment.}

\begin{small}
\bibliographystyle{abbrvnat}  
\bibliography{biblio}
\end{small}

\end{document}